\documentclass[amsfonts,amssymb,amsmath]{revtex4}

\usepackage{times}
\usepackage{graphicx}
\usepackage{bm}
\usepackage{color}

\definecolor{pink}{rgb}{1,1,0} 
\definecolor{red}{rgb}{1,0,0}
\definecolor{yellow}{rgb}{1,1,0}
\definecolor{orange}{rgb}{1,0.5,0}
\definecolor{purple}{rgb}{1,0.5,1}
\definecolor{white}{rgb}{1,1,1}

\newcommand{\be}{\begin{equation}}
\newcommand{\ee}{\end{equation}}

\begin{document}

\title{Induced interaction and crystallization of self-localized impurity fields in a Bose-Einstein condensate}

\author{Sergio Rica$^{1,2}$ and David C. Roberts$^3$}  

\affiliation{$^1$ Laboratoire de Physique Statistique, Ecole Normale Sup\'erieure, UPMC  
Paris 06, Universit\'e Paris Diderot, CNRS, 24 rue Lhomond, 75005 Paris,  France.\\
$^2$Facultad de Ingenier\'ia y Ciencias, Universidad Adolfo Ib\'a\~nez, Avda. Diagonal las Torres 2640, Pe\~nalol\'en, Santiago, Chile.
\\ $^3$ Theoretical Division and Center for Nonlinear Studies, Los Alamos National Laboratory, Los Alamos, NM, USA  
} 

\begin{abstract} 
We model the behavior of $N$ classical impurity fields immersed in a
larger Bose-Einstein condensate by N+1 coupled nonlinear Schr\"odinger
equations in 1, 2, and 3 space dimensions. We discuss the stability of the
uniform miscible system and show the importance of surface tension for
self localization of the impurity fields. We derive analytically the
attractive tail of impurity-impurity interaction due to mediation by
the underlying condensate.  Assuming all impurity fields interact with the
same strength, we explore numerically the resulting phase diagram,
which contains four phases: {\it I}) all fields are miscible;  {\it II}) the
impurity fields are miscible with each other but phase separate from the
condensate as a single bubble;   {\it III}) the localized
impurity fields stay miscible with the condensate, but not with each other; and {\it IV}) the impurity fields phase separate from
the condensate and each other, forming a crystalline structure within
a bubble.   Thus, we show that a crystal can be constructed solely from superfluid components.
Finally, we argue that the crystalline phases maintain their
superfluid behavior, i.e. they possess a nonclassical rotational
inertia, which, combined with lattice order, is a characteristic of supersolidity.
 
\end{abstract}

\date{\today}

\maketitle

\section{Introduction}
 Bose-Einstein condensates (BECs) in trapped atomic gases \cite{becbook1,becbook2},
being highly manipulable and well approximated by a classical
nonlinear mean field theory, have proven to be ideal systems in which
to realize many manifestations of nonlinear physics, such as bright
and dark solitons, quantized vortices and  their lattice formation,
modulation instabilities, and so forth (see \cite{nonlinearbook} for a
review).  In this paper, we use these same qualities of dilute BECs to
explore other aspects of nonlinear physics,  namely the
self localization, phase separation, and crystallization of impurity
fields embedded within a larger condensate.

Specifically, we investigate the formation of nontrivial impurity
structures, such as a crystal composed solely of superfluid components, by modeling the impurity fields
and condensate by coupled nonlinear Schr\"odinger equations (NLSEs).
We are therefore dealing with a type of multicomponent condensate
mixture | a system that has long generated much interest (see for example chapters 15 and 16 of \cite{nonlinearbook} and references therein) | composed of one large component and many smaller
components that we will refer to as impurity fields.  The NLSEs,
acting as a nontrivial yet often tractable model of an idealized
superfluid system, have a long and distinguished history of providing
insights into the fundamental nature of superfluidity.  Indeed, when
dilute Bose-Einstein condensates in trapped atomic gases were
discovered, the NLSEs were often used
successfully to describe quantitively these new superfluid systems.

The analysis in this paper provides the first step toward a
theoretical understanding of realistic experiments where
distinguishable coherent impurity fields in trapped gases might be
achieved by employing different atomic levels, isotopes, species of
atoms, or some combination thereof.  This analysis builds upon the
rich history of investigating impurities in Bose fluids, which have
been widely used not only for probing the properties of Bose fluids,
but also for creating new phases of matter (see for example
\cite{impurity1,impurity2,impurity3,impurity4,impurity5,impurity6,impurity7,eddy}).
 Some of the results presented in this paper were first reported in
\cite{icprl}.

In the following section, we introduce the model considered in this
paper | specifically, $N+1$ coupled NLSEs | and point out the relevant
conserved laws.  The criteria for the system to collapse and the
instability criteria of uniform miscible state are put forward in
section III.  With these established, we proceed to study the
existence of a single localized structure as a solution of (1+1)
coupled NLSEs in section~\ref{SectSingle}.  Here, we outline
variational arguments for the existence of this solution for various
dimensions, examine the importance of surface tension, and compare it
with direct numerical simulations. In section \ref{ImpInteraction} we
develop a perturbative expansion to derive the condensate-mediated
interaction among impurity fields. Section \ref{Crystal} presents the
different phases that arise in this system and we show that
crystallization of the impurity fields is possible in two different
regimes.  We then focus on the case of an impurity crystal forming
within the bubble immersed in a condensate in section \ref{LambaEq0}.
Finally, in section~\ref{SecNCRI} we show that the condensate field
displays nonclassical rotational inertia in the crystal regime and, as
such, bears some of the primary features of supersolidity, namely
lattice ordering and nonclassical rotational inertia.

\section{Model}
\label{nls.model}

In this paper, we consider a large Bose-Einstein condensate denoted by $\psi$ coupled to $N$ small distinguishable impurity fields (treated as classical fields) denoted by $\chi_k$.  We make the assumption that all impurity fields interact with the same coupling constants, which allows for an uncluttered description of the system's nontrivial properties.  Our system of $N+1$ coupled classical fields is therefore governed by the following Hamiltonian: 
\begin{eqnarray}
\label{energy}
H = \int \left( \frac{1}{2} | \nabla \psi |^2 + \frac{1}{2}  |\psi|^4 + \lambda |\psi |^2   \sum_{k=1}^{N} | \chi_k |^2   +  \sum_{k=1}^{N}  \frac{1}{2m} | \nabla \chi_k |^2 + \frac{\gamma_0}{2}  \sum_{k=1}^{N} | \chi_k |^4 + \gamma  \sum_{j<k}^{N} | \chi_j |^2| \chi_k |^2   \right)  d {\bm x}
\end{eqnarray}
$\gamma_0$ is the self interaction of the impurity fields, $\gamma$ is the interaction between impurity fields, and $\lambda$ is the coupling of the condensate to the impurity fields. Unless otherwise stated, we shall restrict ourselves  to positive parameters. In ultracold dilute atomic condensates these coupling constants are directly proportional to the atomic scattering length with a proportionality constant of $4 \pi \hbar^2/m$.   As mentioned above, in principle, these scattering lengths can be tuned in current ultracold trapped atomic gas experiments.

Using this Hamiltonian, the dynamics of the system are governed by  $N+1$ coupled NLSEs (in the condensate context the equations are often referred to as Gross-Pitaevskii equations \cite{becbook1,becbook2}):
\begin{eqnarray}
i \partial_t \psi \, &=& \, -\frac{1}{2} \Delta \psi \, + \,  |\psi|^2 \psi +\lambda \psi  \sum_{k=1}^{N} | \chi_k |^2  \label{nls}\\
i \partial_t \chi_k\, &=& \, -\frac{1}{2m} \Delta \chi_k \, + \, \gamma_0  |\chi_k|^2 \chi_k + \gamma \chi_k  \sum_{j\neq k}^{N} | \chi_j |^2 +  \lambda | \psi|^2  \chi_k \label{nls.imp}
\end{eqnarray}
where $\Delta$ stands for the Laplace operator in $D$ spatial dimensions.  In this system there is conservation of the mass of the large condensate field (particle number) $N^{(c)} = \int  |\psi|^2  d^D{\bm x} $ and that of each impurity field $n_k = \int  |\chi_k|^2  d^D{\bm x} $, and we assume that $n_k \ll N^{(c)} $.  The total energy eq. (\ref{energy}) and the total linear momentum $$ {\bm P } = {\rm Im} \int  \psi^*{\bm \nabla} \psi \, d^D {\bm x} +  {\rm Im} \sum_{k=1}^N \frac{1}{m}  \int  \chi_k^*{\bm \nabla} \chi_k \, d^D {\bm x} $$ of the system are also conserved.  Note that the momentum of each individual field is not conserved.  

\section{Collapse of the system and instabilities of the uniform state}\label{InstabilityCriteriaMatrixM}

In this paper, we are interested in the nontrivial structures that impurity fields can generate in a condensate.  To this end, first note that there are two regimes to avoid.  One is the finite-time collapse of the system, and the other is the uninteresting uniform miscible state.  To find out when these undesirable states occur, consider the $(N+1)\times(N+1)$ interaction matrix 
\begin{equation}
{\mathcal M} = \left( \begin{array}{cccccc}
1&\lambda & \lambda & \cdots &  \lambda & \lambda \\
\lambda & \gamma_0  &\gamma  & \cdots &  \gamma & \gamma \\
\lambda & \gamma  &\gamma_0  & \ddots &  & \gamma \\
\vdots & \vdots & \ddots & \ddots& \ddots& \vdots \\
\lambda & \vdots &   &  \ddots & \gamma_0& \gamma \\
\lambda & \gamma& \cdots  & \cdots & \gamma &  \gamma_0 \\
			\end{array}\right)
			\label{MatrixM}
\end{equation}
derived from the quadratic form of the potential energy, i.e. $\frac{1}{2} {\bm \rho}^{\, \,T} {\mathcal M} {\bm \rho}$ where the vector of the densities is given by ${\bm \rho}= (|\psi|^2, |\chi_1|^2, |\chi_2|^2,...)$.  
One can show that if $\mathcal M$ is negative semidefinite the system will experience a finite-time collapse \cite{robertsnewell}.  Furthermore, because the density is non-negative, one can extend the criteria in \cite{robertsnewell} and  show that the system would also experience a finite-time collapse if  $\mathcal M$ is conegative or positive subdefinite defined as a $\sum_{i,k} {\mathcal M }_{ik} \, x_ix_k \, <0$ for all $ x_i >0$   (for a discussion on the properties of these matrices see \cite{martos}).  We avoid situations that are proven to exhibit finite-time collapse by our assumption that the coupling constants are positive.

Next, by ignoring kinetic energy terms (which means ignoring surface tension and effects due to the system being of finite size), one can prove that the uniform state is energetically stable if and only if $\mathcal M$ is positive semidefinite, i.e. if all eigenvalues of $\mathcal M$ are non-negative \cite{robertsueda}.  It is when a modulational instability arises that the nontrivial structures emerge.  We will therefore focus on situations where at least one eigenvalue of $\mathcal M$ is negative. 

The $N+1$ eigenvalues of $\mathcal M$ are $\gamma_0-\gamma$, which is $N-1$ degenerate, and  
$ \frac{1}{2} \left( (1 +(N-1) \gamma + \gamma_0)  \pm \sqrt{  (-1 + (N-1) \gamma + \gamma_0) ^2 + 4 N \lambda^2} \right). $  Therefore, when $\gamma>\gamma_0$ or $\lambda > \frac{{\sqrt{(N-1)\,\gamma  +{\gamma_0}}}}{{\sqrt{N}}}$, the system becomes energetically unstable.  This highlights the difference between the two types of instabilities in this system, which is important for the analysis in this paper.  In the first regime, i.e. $\gamma>\gamma_0$, each of the impurity fields localizes with respect to the other impurity fields regardless of whether or not they are phase separated from the condensate.  In the second regime, i.e. $\lambda > \frac{{\sqrt{(N-1)\,\gamma  +{\gamma_0}}}}{{\sqrt{N}}}$, the large background condensate phase separates from the impurity fields.  For a discussion of the dynamical instabilities of the miscible state, see Appendix \ref{AppA}.

\section{Self localization of a Single Impurity Field}
\label{SectSingle}

In this section, we will discuss the system of a single small  impurity field embedded in and interacting with a large condensate.  We will show, by a variational argument,  that there is a critical coupling parameter between the condensate and impurity beyond which the impurity field self localizes.  By self localization we mean the ground state of the system will contain an impurity field with a finite extent (or with an extent smaller than the size of the system) due to the interaction between the impurity field and the surrounding condensate.  We will show that surface tension allows one to interpolate between the previous known results of self localization of a single impurity atom \cite{eddy} and the bulk phase separation condition \cite{colsonfetter}.

We estimate the ground-state wave functions via a variational argument for the system with a single impurity field in 1, 2, and 3 dimensions. We impose that the trial wave functions  possess the following characteristics: that $\chi (r)$ be localized such that $\chi (r) \rightarrow 0 $ as $r  \rightarrow \infty$, and its particle number be fixed such that $n_* = C_D  \int  |\chi|^2  r^{D-1}  d r   ,$ where  $C_D = \frac{2\,{\pi }^{\frac{D}{2}}}  {{\Gamma}(\frac{D}{2})} $ is the surface of a unitary sphere in $D$ spatial dimensions.  The condensate wave function should be depleted where the impurity field is positioned, and $ \psi (r)  \rightarrow \psi_0 = constant$ as $r  \rightarrow \infty$.  (Note $\psi_0$ is real because it is a ground state.)  The appropriate energy to be minimized is
\begin{equation}
E = C_D\int \left( \frac{1}{2} | \nabla \psi |^2 + \frac{1}{2}  (|\psi|^2 - | \psi_0 |^2 )^2 + \lambda (|\psi|^2 - | \psi_0 |^2 )  | \chi |^2   + \frac{1}{2m} | \nabla \chi |^2 +\frac{\gamma_0}{2}  | \chi |^4 \right) r^{D-1}  d {r}\label{free}.
\end{equation}

We will consider the following real, normalized trial functions:
\begin{eqnarray}
 \chi(r)  &=& \sqrt{n_*} \sqrt{\frac{\alpha^D} {C_D \, {\mathcal N}_D} }
f({\alpha }\, r) \label{trial1}\\
 \psi (r) & = & \psi_0\left( 1 - a \,  \chi (r) ^2  \right),  \label{trial2}
\end{eqnarray}
with variational parameters $a$ and $\alpha > 0 $, both determined once the energy estimates are minimized (note that, in a real and finite system, the limit $\alpha \rightarrow 0$ cannot exist); and $f(r)$ is localized, i.e. $f(r) \rightarrow 0$ as $r \rightarrow \infty$.  The normalization constant is ${\mathcal N}_D = \int_0^\infty  f(x)^2 x^{D-1}\,dx $.  Once a formal expansion of the energy is achieved, one can easily demonstrate that $a\approx \frac{\lambda}{2 \psi_0^2}  $ as  $\alpha\rightarrow 0$.

One can then show that there is an upper bound on the $D$-dimensional energy eq. (\ref{free}) of the form
\begin{equation} \label{Ealpha}
E\leq E(\alpha) =
\epsilon_0  \, {\alpha }^2+ 
  \epsilon_1\, {\alpha }^D
   + \epsilon_2  \,{\alpha }^{2 + D}   + 
 \epsilon_3  \,{\alpha }^{3\,D}
 \end{equation}
 where the constants $\epsilon_s$ are given by
\begin{eqnarray}
\epsilon_0 &= & \frac{n_* }{2 m \, {\mathcal N}_D}  \int_0^\infty  f'(x)^2 x^{D-1} \,dx ,  \nonumber\\
\epsilon_1 &= & (\gamma_0-\lambda^2) \frac{n_*^2 }{2 C_D \, {\mathcal N}_D^2}  \int_0^\infty  f(x)^4 x^{D-1} \,dx , \nonumber\\
\epsilon_2 &= &  \frac{n_*^2\, \lambda^2 }{2 C_D \, {\mathcal N}_D^2 \psi_0^2}  \int_0^\infty  f(x)^2 f'(x)^2 x^{D-1} \,dx , \nonumber\\
\epsilon_3 &= & \frac{n_*^4\, \lambda^4 }{32 C_D^3\, {\mathcal N}_D^4\, \psi_0^4}  \int_0^\infty  f(x)^8 x^{D-1} \,dx . \nonumber\\
\end{eqnarray}

It is clear that $\epsilon_0,\, \epsilon_2$, and $\epsilon_3$ are all positive numbers, but $ \epsilon_1 = k_1  \left({\gamma_0} - \lambda^2 \right) n_*^2$ may become negative if $\lambda^2 > \gamma_0$.  For example, for a Gaussian wavefunction $f(x)= e^{- x^2}$, ${\mathcal N}_D =  \frac{\Gamma(D/2) }{2^{ D/2 +1}}$, and one can deduce the following constants: 
\begin{eqnarray} \label{EalphaGaussian}
\epsilon_0 &= & \frac{D\,n}{2\,m} ,  \nonumber\\
\epsilon_1 &= & \frac{n^2}{2\,{\pi }^{\frac{D}{2}}}  \left({\gamma_0} - \lambda^2\right) , \nonumber\\
\epsilon_2 &= & \frac{D\,n^2\,{\lambda }^2} {4\,{\pi }^{\frac{D}{2}}\,{{\psi_0}}^2}  , \nonumber\\
\epsilon_3 &= & \frac{2^{-5+\frac{D}{2}}\,n^4\, {\lambda }^4}{{\pi }^{\frac{3\,D}{2}}\, {{\psi_0}}^4}  . \nonumber\\
\end{eqnarray}

 It should be stressed that the variational approach only provides a ceiling on the ground state of the energy.  If this upper bound | the lowest variational energy | is negative for $\alpha \neq 0$ (since, for the nearly uniform state, the energy is positive and tends to zero as $\alpha \rightarrow 0$), then the uniform state is known to be unstable and it is most probable that the impurity field has self localized.  However, should the upper energy bound be non-negative, one cannot determine from this whether or not self localization is likely to occur.  

Regardless, the energy expression eq. (\ref{Ealpha}) in terms of the variational parameter $\alpha$ does provide some guidance as to when a localized impurity solution is likely to exist in 1, 2, and 3 spatial dimensions.  The key here is the observation that when $\epsilon_1>0$, the function (\ref{Ealpha}) increases monotonically, implying that the energy is minimized when $\alpha \rightarrow 0$.  We will now examine the situation of one, two, and three dimensional space in turn.

\subsection{Self localization in 1D}
Applying this in one dimension is straightforward: since when $D=1$ the dominant term of $E(\alpha)$, i.e. eq. (\ref{Ealpha}),  at small $\alpha$ is $ \epsilon_1\, {\alpha }$, when $\epsilon_1$ is negative, i.e. $\lambda^2>\gamma_0$, a supercritical transition occurs from a homogeneous state to a localized impurity state.  

\subsection{Self localization in 2D}
For $D=2$,  the first and second terms of $E(\alpha)$ in eq. (\ref{Ealpha}) are of the same order. As occurs for $D=1$, there is a second-order transition towards a localized impurity state if $\lambda^2 > \gamma_0 + \frac{k_1}{m n_*} $, with $k_1 = 4 C_2\, {\mathcal N}_2  \frac{\int_0^\infty  f'(x)^2 \, x \,dx  }{\int_0^\infty  f(x)^4 \, x \,dx }$. The instability of the homogeneous state is shifted from the bulk condition ($\lambda>\gamma_0$).  This shift has a simple interpretation: the presence of a $1/m$ factor means that this term comes from the kinetic energy of the impurity $\int \frac{1}{2 m}  |\nabla \chi|^2 $, so the shift is created by the curvature $\alpha$ of the localized structure | it is a kind of surface tension. Furthermore, $\frac{\int_0^\infty  f'(x)^2 \, x \,dx  }{\int_0^\infty  f(x)^4 \, x \,dx }$ bears the hallmark of a surface tension effect: that is the ratio of interface energy (the gradient term) to a bulk energy. 

In the case of the Gaussian trial function one finds that a localized solution exists if $\lambda > \lambda_c =\sqrt{ \gamma_0 + \frac{2 \pi}{m n_*} }$.  
These criteria are close to those numerically observed in two-dimensional space. 
Indeed, we have simulated numerically eqs (\ref{nls}) and (\ref{nls.imp}) for the case of a single impurity ($N=1$) with $\gamma_0=0$, $m=1$, and $n_* = 40.96$ in a $64^2$ box in which the total number of particles in the condensate was $\int |\psi|^2 d^2x = 4096$. Below we present our findings for two different sets of initial states: {\it i)} uniform miscible state, i.e. $\psi \approx 1 $  and $\chi \approx 0.1$, both real (here $\approx C$ means $C$ plus a small fluctuation  or noise); and {\it ii)} $\psi \equiv 1 $, and $\chi$ is set as the Gaussian Ansatz (\ref{trial1}) where the value of $\alpha $ 
is that which minimizes the energy $E(\alpha)$ in 2D. By the variational criteria, a localized solution should exist if $ \lambda > \sqrt{\frac{2 \pi}{m n_*} } \approx 0.3917$.

For the initial condition {\it i)}, we find numerically that if $\lambda < 0.5$ no instability will develop because of the finite size of the system. Indeed, if $\lambda < \frac{\pi}{\sqrt n_*} \approx 0.491 $, the system is not unstable for the longest wavelength allowed by the box (a wavenumber $\frac{2 \pi}{L} $). However, as soon as we get a localized solution,  e.g. $\lambda = 0.5$, it is possible to decrease $\lambda$ to below  $\lambda_c=0.39$ and retain the localization. Indeed, we can bring the wavelength down to as low as $\lambda \approx 0.35$ and still keep the nonlinear solution.  Below $\lambda \approx 0.35$ it is not possible to distinguish a real localized structure from  oscillatory structures. 

For $\lambda = 0.35$ to $0.55$ an oscillatory pulse is observed | a kind of breathing. This is perhaps due to an excess of energy, and the localized solutions must radiate to evacuate it.  The nature of the oscillation should follow from a similar Ansatz to eqs (\ref{trial1},\ref{trial2}) but allow the phase of the impurity field to be a space- and time-dependent function. For instance, taking $ \chi(r,t)  = \sqrt{n_*} \sqrt{\frac{\alpha^D} {C_D \, {\mathcal N}_D} } f({\alpha }\, r) e^{ i p(t) r^2 }$ and introducing this into the Lagrangian of the system (which is $ - \int \frac{i}{2}(\chi \partial_t \bar\chi - \bar\chi \partial_t\chi ) d^D{\bm x}$ minus the energy eq. (\ref{free})), for the Gaussian trial function one arrives at the following Lagrangian: ${\mathcal L} (\alpha, p, \dot p) = - E(\alpha) - n_* \frac{D}{2 \alpha^2} \left( \frac{ p^2}{m}  +  \frac{\dot p}{2}\right)$. The Euler-Lagrange conditions lead to the ``momentum'' equation  $p =-\frac{m \dot \alpha}{2 \alpha} $.  Finally, this yields the dynamics that the ``breathing'' must satisfy: $$\frac{D m n_*}{4 \alpha^5} \left(\alpha \, \ddot\alpha  - 2 \dot\alpha^2\right) =  - \frac{d}{d\alpha} E(\alpha). $$  Thus, near the threshold $\lambda\gtrsim \lambda_c = \sqrt{ \gamma_0 + \frac{k_1}{m n_*} }$, one finds that $\alpha_{eq} \sim \sqrt{\lambda- \lambda_c  }$ and the breathing frequency is $\omega_{breathing}\sim\sqrt{\alpha_{eq}^4E''(\alpha_{eq})} \sim (\lambda- \lambda_c  )^{3/2}$.

 \begin{figure}[htc]
\begin{center}
\centerline{{\it a)} \includegraphics[width=3cm]{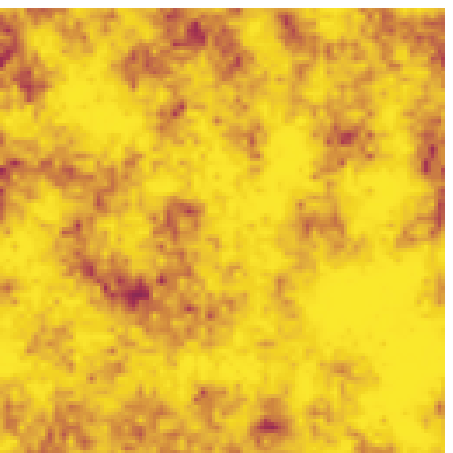} \quad {\it b)} \includegraphics[width=3cm]{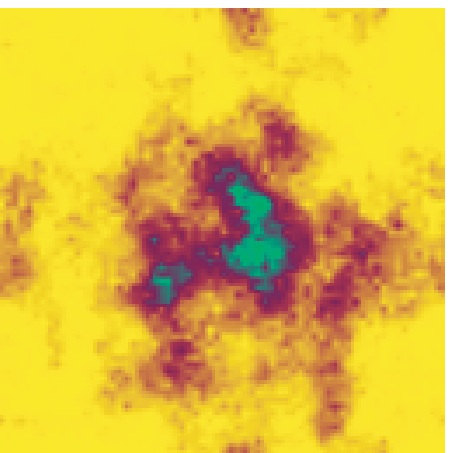} }
\centerline{{\it c)} \includegraphics[width=3cm]{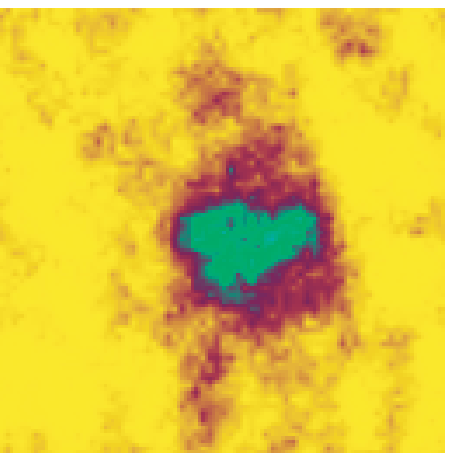} \quad {\it d)} \includegraphics[width=3cm]{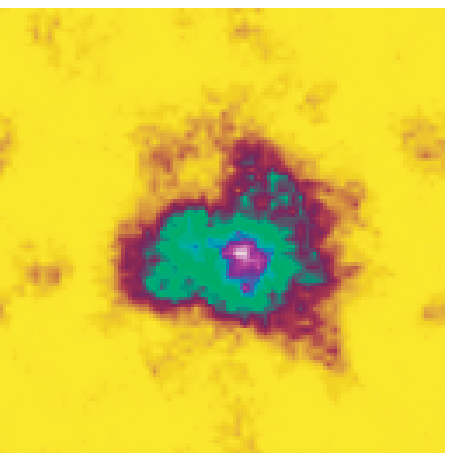} }
\caption{ \label{SingleImpurities} 
(Color online) Numerical simulation (see Appendix B for details) of eqs (\ref{nls}) and (\ref{nls.imp}) for a single impurity $(N=1)$ with $\gamma_0=0$ in a $64 \, \times 64 \, {\rm units^2}$ periodic plane. Here $\int |\psi|^2 d^2{\bm x} = 4096$ and  $\int |\chi_k|^2 d^2{\bm x} = 40.96 $. The plots represent $ |\chi_1|^2 $ and the colormap is the same for all four images. In {\it a)}, $\lambda=0.385$ (no localized solution is observed, but because of the recurrence phenomenon discussed in the text, after a time a kind of pulse is observed, which then dissipates, only to eventually form again); in {\it b)}, $\lambda=0.39$; in {\it c)}, $\lambda=0.395$; and in {\it d)}, $\lambda=0.4$.  All these solutions are dynamically obtained from the same initial state {\it i)} and, as described in the main text.  We start with $\lambda=0.5$ and, after the spontaneous formation of a localized solution, we decrease $\lambda$ slowly, i.e. $\lambda$ varies in time every 1000 time units by $\lambda \rightarrow \lambda - \Delta\lambda$ where  $\Delta\lambda = 0.005$.}
\end{center}
\end{figure}

For the initial condition {\it ii)}, valid only if $\lambda>\lambda_c$,  one notices that the Gaussian trial function is very close to the exact stationary solution, and the pulsations and breathing described previously are less marked. From an initial condition of $\lambda>\lambda_c$, we can decrease $\lambda$ to below $\lambda_c$, retaining the localization. The oscillations, however, become very important and, as earlier, it seems that rather than a stationary state, a more complex oscillatory behavior is displayed. This dynamic solution seems to be very robust as we can observe it even for very low $\lambda$. The pulse forms, disperses through the system, and then after some (possibly long) time the pulse forms again, in a kind of recursive, and not necessarily periodic dynamic.  However, this dynamic behavior dominated by pulses seems to be quite different from the steady ground state sketched by our variational theory. 

Nevertheless, we notice that the numerically observed range over which steady-state localized solutions exist is very close to that expected via the variational analysis. The difference probably arises because we are simulating the Hamiltonian evolutions eqs (\ref{nls}) and (\ref{nls.imp}), and not simply finding the minimum energy of the Hamiltonian eq. (\ref{energy}).


 
\subsection{Self localization in 3D}

For $D=3$,  finding the self-localization conditions is more subtle than in lower dimensions.   As discussed above, the variational argument shows that a localized solution exists if the energy $E$ in eq. (\ref{Ealpha}) is negative for some critical value, $\alpha_c$.  A necessary (but not sufficient) condition for a negative-energy  ground state in 3 D  is $\epsilon_1<0$, i.e.  $\lambda^2 > \gamma_0$.  

We can make a more precise estimate as follows: Because the energy turns negative for $\alpha >0$, and because in general the energy expansion grows as $\alpha^2$ near $\alpha = 0$ (since $\epsilon_0>0$ in eq. (\ref{Ealpha})), it is sufficient (though not necessary) that $E(\alpha)$ have a minimum at  $\alpha = \alpha_c$ and that this minimum reach the horizontal axis $E(\alpha_c)=0$. In equations, this reads as $E(\alpha_c)=0$ and $E'(\alpha_c)=0$, which describe a critical line in spatial parameters, separating the region where a localized structure (with great probability) exists from that where there is no such certainty.  This line may be written  parametrically as 
\begin{equation}
 6 \beta ^7+2 \beta ^3=\frac{\epsilon_0 \epsilon_3^{3/4}}{\epsilon_2^{7/4}}\quad \&\quad 7 \beta ^6+3 \beta ^2+\frac{\epsilon_1\sqrt{\epsilon_3}}{\epsilon_2^{3/2}}=0,
 \label{critical3D}
\end{equation}
where $\beta \equiv \alpha_c \left({\epsilon_3}/{\epsilon_2}    \right)^{{1}/{4}}$ is the parametrization.  The parametric curve may be approximated in the large and small $\beta $ limit respectively by:
\begin{eqnarray}
 \epsilon_1 &= & -\frac{7 \, \epsilon_0^{6/7}\,  \epsilon_3^{1/7} }{6^{6/7}} \quad {\rm if}\,  \beta \gg 1 ,   \label{asymplarge} \\
 \epsilon_1 &= &   -\frac{3 \, \epsilon_0^{2/3}\,  \epsilon_2^{1/3} }{2^{2/3}}  \quad {\rm if}\,  \beta \ll 1.  \label{asympsmall}
\end{eqnarray}

\begin{figure}[htc]
\begin{center}
\centerline{\includegraphics[width=8cm]{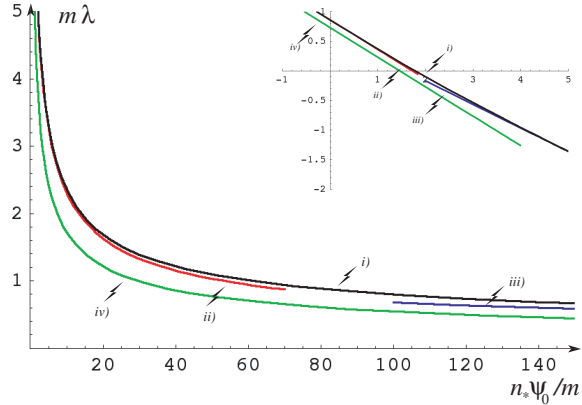}  }
\caption{ \label{3DCriteriaForLocalizedStr}  (Color online) Parametric plot of the critical transition line for $\gamma_0=0$ in three spatial dimensions. The curve  {\it i)} is the parametric curve (\ref{critical3D}); the  curve {\it ii)}  is the asymptotic behavior eq. (\ref{asymplarge2}) for  $ n_*\psi_0/m\ll 1$; and {\it iii)} is the asymptotic behavior eq. (\ref{asympsmall2}) for $ n_*\psi_0/m\gg 1$. The curve {\it iv)}   represents the Cucchietti--Timmermans condition  \cite{eddy}: $ \frac{1}{2 \pi} \lambda^2 n_* \psi_0 m \geq 4.7$, that is  $ \lambda_{CT} = \frac{5.43}{\sqrt{n_* \psi_0 m }}$. Finally, the inset represents the same curves, but in log-log to cover a larger range of values. }
\end{center}
\end{figure}

Let us look briefly at the case $\gamma_0=0$.  Here, one may express eq. (\ref{critical3D}) as a closed parametric expression for the variables $(m\, \lambda \, , \, \frac{\psi_0 \, n_*}{m}  \,)$. The values of the $\epsilon$'s in eq. (\ref{EalphaGaussian}) for the Gaussian trial function yield a bound that is about 35\% higher than the numerical result \cite{eddy}, as plotted in Fig. \ref{3DCriteriaForLocalizedStr}.  Moreover, for a Gaussian trial function the asymptotic formulas (\ref{asymplarge}) and  (\ref{asympsmall}) take on the simplified expressions
\begin{eqnarray}
 m \, \lambda &= & \left( \frac{9 \pi}{2}\right)^{3/4}    \frac{1}{\sqrt{ n_* \psi_0/m} }  =  7.2907 \frac{1}{\sqrt{ n_* \psi_0/m} } \quad {\rm for}\,  (n_* \psi_0/m) \ll 1 ,   \label{asymplarge2} \\
m \, \lambda  &= &   \frac{7^{7/10} \pi^{3/5}}{2^{17/20}}   \frac{1}{(n_* \psi_0/m) ^{2/5}} = 4.3051  \frac{1}{(n_* \psi_0/m) ^{2/5}}   \quad {\rm for}\,  (n_* \psi_0/m)\gg 1. \label{asympsmall2}
\end{eqnarray}

\section{Mediated attraction of impurity fields due to the condensate}\label{ImpInteraction} 

In this section, we look at how $N$ localized dilute impurity fields interact and determine the role of the modified background condensate in this interaction.  We will demonstrate that the interaction between impurity fields has an attractive tail, and that this is a result of mediation by the condensate, combined with the hard-core repulsion arising from the positive scattering length between impurity fields.   

Consider $N$ impurity fields that are self localized (having satisfied the conditions given in the previous section) and that weakly modify the initially uniform condensate, i.e.
\begin{equation}
\psi = \psi_0 +\epsilon \psi_1(r),
\end{equation}  
with $\psi_0$ assumed constant.  This assumption holds if 1) there is only weak localization of the impurity or 2) the impurity is distant enough.  The second case is particularly relevant here as we intend to derive the nature of the interaction at long range. 
In this approximation, using eq. (\ref{nls}) the condensate wave function is given by (here $| \chi_k |^2 $ is assumed to be of order $\epsilon$)
\begin{equation}-\frac{1}{2} \Delta \psi_1 \, + \,  2 \psi_0^2 \psi_1 +\lambda \psi_0  \sum_{k=1}^{N} | \chi_k |^2  =0.
\label{linearcondenstae}
\end{equation}
This linear equation (known as the screened Poisson equation) may be solved with the aid of a Green's function in $D$-spatial dimensions, 
$$ - \Delta G^{(D)}({\bm x}-{\bm x}')  \, + \,  (2 \psi_0)^2 G^{(D)}({\bm x}-{\bm x}')  = C_D \delta^{(D)}({\bm x}-{\bm x}'),$$
where $C_D$ is the surface of a unit sphere.  The solution can be written explicitly as
\begin{equation}G^{(D)}({\bm x}-{\bm x}')  = \left\{ 
		\begin{array}{ll}
 \frac{1}{4 \psi_0} e^{-2\psi_0 | {\bm x}-{\bm x}'|} & \quad {\rm} in \, D=1 \\
  K_0(2\psi_0 | {\bm x}-{\bm x}'|) & \quad {\rm} in \, D=2 \\
    \frac{e^{-2\psi_0 | {\bm x}-{\bm x}'|}}{ | {\bm x}-{\bm x}'|} & \quad {\rm} in \, D=3\\
				\end{array}
\right.
\label{yukawa}
\end{equation}
where $K_0$ is the modified Bessel function.  

Solving eq. (\ref{linearcondenstae}) yields the following modification (which is of order $\epsilon$) to the condensate field:
\begin{equation} \label{psi1}
\psi_1({\bm x})  = - \frac{2 \lambda \psi_0}{C_D} \int G^{(D)}({\bm x}-{\bm x}')   \sum_{k=1}^{N} | \chi_k ({\bm x}')|^2 d^{D}{\bm x}'. 
\end{equation}
Under the assumption of a weakly modified condensate, the total energy (on the order of $\epsilon^2$) is given by 
\begin{eqnarray}
E= \int \left( \frac{1}{2} | \nabla \psi_1 |^2 + 2 \psi_0^2  |\psi_1|^2 + 2  \lambda \psi_0 \psi_1  \sum_{k=1}^{N} | \chi_k |^2   +  \sum_{k=1}^{N}  \frac{1}{2m} | \nabla \chi_k |^2 + \frac{\gamma_0}{2}  \sum_{k=1}^{N} | \chi_k |^4 + \gamma  \sum_{j<k}^{N} | \chi_j |^2| \chi_k |^2   \right)  d {\bm x}.\label{interatingenergy}
\end{eqnarray}
 The consistent scaling is achieved by assuming $ | \nabla \chi_k |^2\sim \epsilon^2$ (which plays a role in the self energy as discussed below) and $  | \chi_k |^2 \sim \epsilon$. 
 
One can multiply eq. (\ref{linearcondenstae}) by $\psi_1$ and integrate  over the total volume to obtain
$$\int \left( \frac{1}{2} | \nabla \psi_1 |^2 + 2 \psi_0^2  |\psi_1|^2 +   \lambda \psi_0 \psi_1  \sum_{k=1}^{N} | \chi_k |^2  \right)  d {\bm x}  =0.$$
 Using this identity we can eliminate the terms  in eq. (\ref{interatingenergy}) that do not depend explicitly on the impurity fields  to arrive at
\begin{equation}
E  =   \int \left(   \lambda \psi_0 \psi_1  \sum_{k=1}^{N} | \chi_k |^2   +  \sum_{k=1}^{N}  \frac{1}{2m} | \nabla \chi_k |^2 + \frac{\gamma_0}{2}  \sum_{k=1}^{N} | \chi_k |^4 + \gamma  \sum_{j<k}^{N} | \chi_j |^2| \chi_k |^2   \right)  d {\bm x}.
\end{equation}
Utilizing  the solution for $\psi_1(r)$, i.e. eq. (\ref{psi1}), the energy of the system of $N$ impurity fields under the assumption of a weakly modified condensate can be written as
\begin{eqnarray}
E &= &   
 \sum_{i \neq k}^{N}   \int \left( \frac{ \gamma}{2}  \delta^{(D)}({\bm x}-{\bm x}') -  \frac{2 \lambda^2 \psi_0^2 }{C_D}   G^{(D)}({\bm x}-{\bm x}')  \right)    | \chi_i ({\bm x}')|^2  | \chi_k ({\bm x})|^2 \,d^{D}{\bm x}\,  d^{D}{\bm x} '      + E_0\label{interatingenergy2}
\end{eqnarray}
where the self-interaction energy of the impurity fields is given by 
\begin{equation}
E_0 =   \sum_{k=1}^{N}  \left[  \int \left(   \frac{1}{2m} | \nabla \chi_k |^2 + \frac{\gamma_0}{2}  | \chi_k |^4  \right)  d {\bm x} -      \frac{2 \lambda^2 \psi_0^2 }{C_D}  \int   G^{(D)}({\bm x}-{\bm x}')    | \chi_k ({\bm x}')|^2  | \chi_k ({\bm x})|^2  d {\bm x} \, d {\bm x}' \right]. 
\label{selfenergy}
\end{equation}
(This self-interaction term $E_0$ was previously derived in \cite{eddy}.)

Let us consider localized impurity fields, $\alpha|x_k -x_i|  \gg 1$. One may approximate the $k$-th impurity field by $$ | \chi_k ({\bm x})|^2= n_k \delta^{(D)}({\bm x}-{\bm x}_k)$$
where ${\bm x}_k$ is the position of the $k$-th impurity.  In  this case $E=  \frac{1}{2}\sum_{i\neq k}^{N}  U(|{\bm x}_i-{\bm x}_k|) n_i n_k$  where the interaction potential between the two impurity fields is given by 
\begin{equation}
U(|{\bm x}_i-{\bm x}_k|)= \gamma \delta^{(D)}({\bm x}_i-{\bm x}_k) -   \frac{4 \lambda^2 \psi_0^2 }{C_D}   G^{(D)}({\bm x}_i-{\bm x}_k).  
\label{interactingenergy3}
\end{equation}
Note that the self-interaction energy $E_0$ naturally diverges because of the artificial singular structure { $ | \chi_k ({\bm x})|^2= n_k \delta^{(D)}({\bm x}-{\bm x}_k)$ assumed in the impurity field.

In eq. (\ref{interactingenergy3}), the first term on the RHS is, on its own, merely a crude estimate. The full energy expression eq. (\ref{interactingenergy3}) points to the existence of a bound state but with an equilibrium distance of zero. In realistic systems, the equilibrium distance is dependent on the impurity's localization, which is on the order of the size of the structure. 

For example, consider the trial function eq. (\ref{trial1}) of the form $  \chi_k ({\bm x})=  \sqrt{n_k} \sqrt{\frac{\alpha^D} {C_D \, {\mathcal N}_D} } e^{-\alpha^2|{\bm x}-{\bm x}_k|^2}$. Introducing this into the interaction energy eq. (\ref{interatingenergy2}), one arrives (in 3D) at
\begin{equation}
U(|{\bm x}_i-{\bm x}_k|)= \gamma \frac{\alpha^3}{\pi^{3/2}} e^{-\alpha^2|{\bm x}_i-{\bm x}_k|^2} -   \frac{ \lambda^2 \psi_0^2 }{ \pi}     \frac{e^{-2\psi_0 |{\bm x}_i-{\bm x}_k| }}{ |{\bm x}_i-{\bm x}_k|} . 
\label{interactingenergy4}
\end{equation}
The second term is estimated in the limit $\alpha \gg 2 \psi_0$, i.e. where the structure is more localized than the Yukawa interaction range of $1/(2\psi_0)$. The equilibrium distance comes from this energy.
 

\section{N impurity fields - Crystallization}\label{Crystal}

Here we will discuss a system composed of $N$ interacting impurity fields inside a  larger condensate.  In section \ref{ImpInteraction}, we showed that over a certain range of parameters the tunable interaction between localized impurity fields has a hard core and an attractive tail, which is a type of interaction that has been studied in the context of many diverse physical systems (for Yukawa-type attractive interactions see e.g.  \cite{yukawa}).  We show below that, depending on the interaction between condensate and impurity, there are two possible regimes in which such impurity fields crystallize: the first where the impurity fields remain immersed in the condensate, and the second where they phase separate in the form of a crystal within a bubble within the condensate.  

The phase diagram in Fig. \ref{SimpleDiagram} shows four distinct regimes.  In phase {\it I}, the impurity fields are miscible with each other and with the condensate.  In phase  {\it II}, the impurity fields are miscible with each other but phase separate as a bubble from the condensate.  In phase  {\it III}, localized impurity fields stay miscible within the condensate and can crystallize if system is not dilute; in phase {\it IV} the impurity fields phase separate from the condensate and form a crystalline structure within a bubble

In an infinite system, the critical lines for the equilibrium phase are $\gamma=\gamma_0$ and $\lambda = \frac{{\sqrt{(N-1)\,\gamma  +{\gamma_0}}}}{{\sqrt{N}}}$ as discussed in section \ref{InstabilityCriteriaMatrixM}.  
In the following subsections we shall discuss the different phases.  Although in a finite system these transition lines are shifted because the lowest mode (where the instability typically arises) is nonzero, for  simplicity, the transition lines discussed will be those for the case of an infinite system. However, it should be kept in mind that corrections maybe  computed (See Appendix A for discussion).

\begin{figure}[htc]
\begin{center}
\centerline{ \includegraphics[width=10cm]{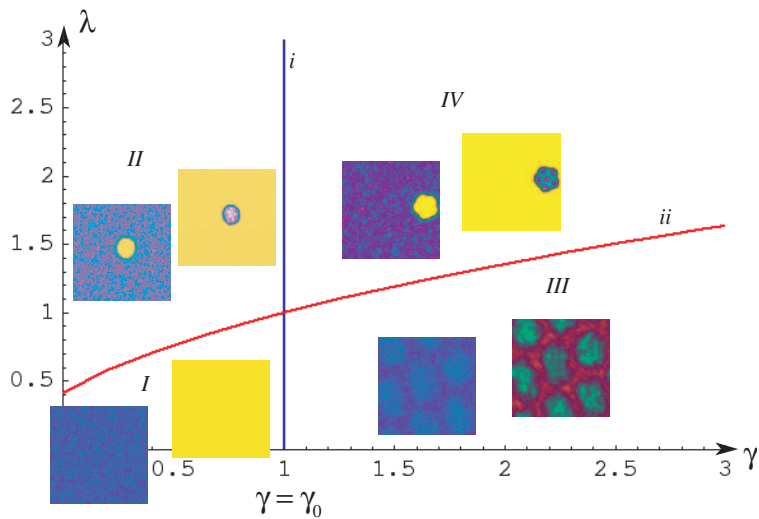}   }
\caption{ \label{SimpleDiagram} 
(Color online)  Phase diagram for $N=6$ impurity fields with $\gamma_0 =1$. The vertical line {\it i)} represents the border line  $\gamma=\gamma_0=1$ while the curve {\it ii)} represents the critical line $\frac{\sqrt{(N-1)\,\gamma  +\gamma_0} }{\sqrt{N}}$. 
The accompanying pairs of pictures (of which the left represents the condensate density, $|\psi|^2$, and the right plots $\sum_{k=1}^N |\chi_k|^2 $) are numerical simulations in a $64 \, \times 64 \, {\rm units^2}$ periodic plane with uniform (plus a small fluctuation) initial conditions for the condensate and the impurity fields such that $\int |\psi|^2 d^2{\bm x} = 4096$ and  $\int |\chi_k|^2 d^2{\bm x} = 40.96 $. Explicitly, we have $\lambda = 0.25$ and $ \gamma = 0.5$ in $I$; $\lambda = 2$ and $ \gamma = 0.5$ in $II$; $\lambda = 0.25$ and $ \gamma = 2$ in $III$; and $\lambda = 2$ and $ \gamma = 2$ in $IV$.  See Appendix B for details of numerical tools.  
 }
\end{center}
\end{figure}

\subsection{Phases {\it I} \& {\it II}}
In the case  $\gamma<\gamma_0$ and  $\lambda <  \frac{{\sqrt{(N-1)\,\gamma  +{\gamma_0}}}}{{\sqrt{N}}}$, the condensate-impurities system finds its minimum energy in a homogeneous state, which is the completely {\it miscible} state. This is phase {\it I} of Fig. \ref{SimpleDiagram}. In phase {\it II},  $\gamma<\gamma_0$ and  $\lambda >  \frac{{\sqrt{(N-1)\,\gamma  +{\gamma_0}}}}{{\sqrt{N}}}$, and the condensate phase separates from the impurity fields, which remain together and miscible with each other in a bubble.  No crystallization occurs in either phase.

\subsection{Phase {\it III}}
In this case, $\gamma>\gamma_0$ and $\lambda <  \frac{{\sqrt{(N-1)\,\gamma  +{\gamma_0}}}}{{\sqrt{N}}}$.
In phase {\it III}, the interaction energy eq. (\ref{interatingenergy2}) causes the impurity fields to attract.  However, because a repulsive hard core exists at close range, one expects there to be an equilibrium distance. Although we expect the system to crystallize, the complete picture is not so simple because there is also a density parameter that plays a fundamental role. Indeed, as already seen in section \ref{SectSingle}  $\alpha^{-1}$ is roughly the size of a localized structure (see section \ref{SectSingle}), so then the ``diluteness'' of the impurities in the condensate is measured by $\frac{ N}{(\alpha L)^D}$. If this parameter is close to unity, as is true in typical solids, one gets a crystalline phase  (see Fig. \ref{figyukawa}). However, if this phase's ``diluteness'' becomes small, one would expect to see liquid- or gas-like behavior, as occurs in molecular dynamics simulations \cite{yukawa}. These states are roughly observed in numerical simulations (see Fig. \ref{YukawaPhases}).

\begin{figure}[htc]
\begin{center}
\centerline{{\it a)} \includegraphics[width=3.5cm]{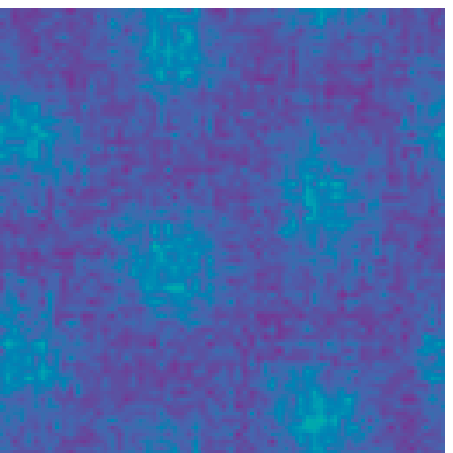} \quad {\it i)}  \includegraphics[width=3.5cm]{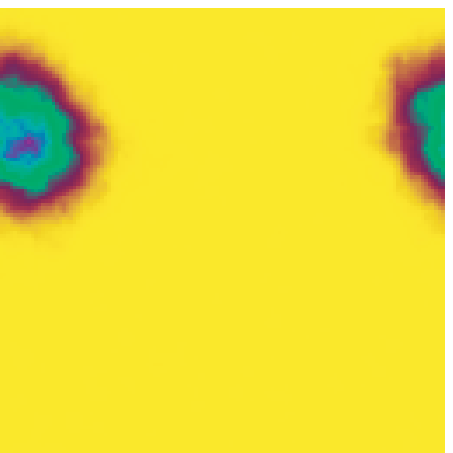}  \quad {\it ii)}  \includegraphics[width=3.5cm]{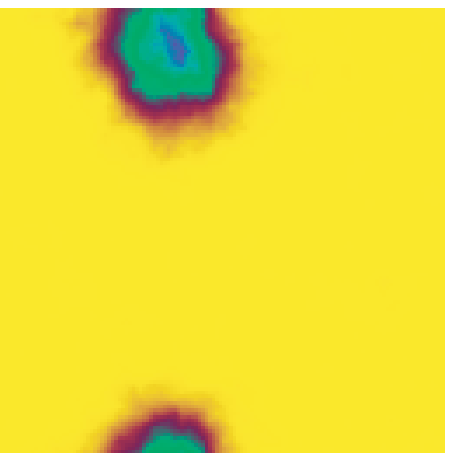}  \quad {\it iii)}  \includegraphics[width=3.5cm]{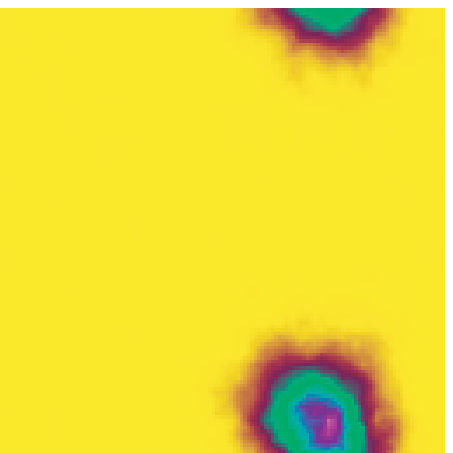} }
\centerline{{\it b)} \includegraphics[width=3.5cm]{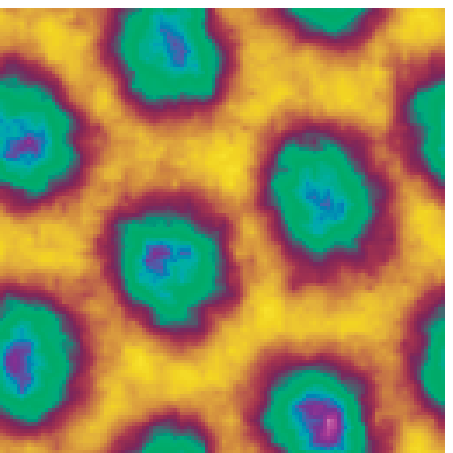} \quad {\it iv)}  \includegraphics[width=3.5cm]{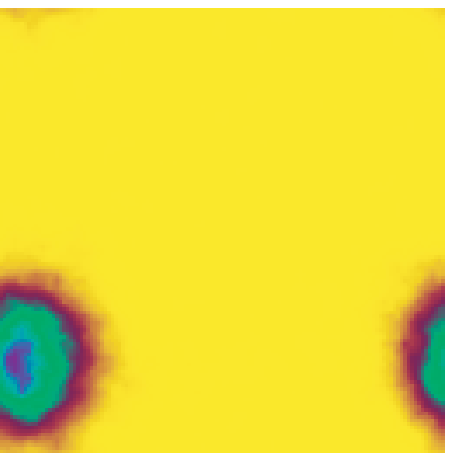}  \quad {\it v)}  \includegraphics[width=3.5cm]{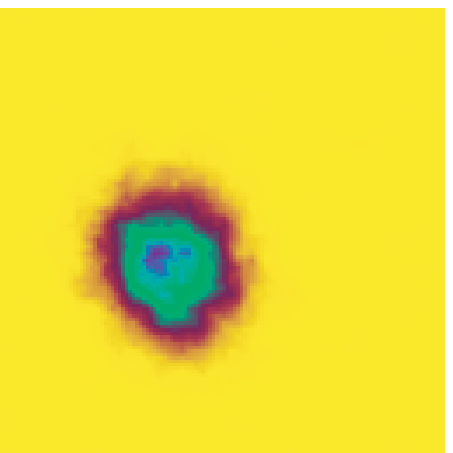}  \quad {\it vi)}  \includegraphics[width=3.5cm]{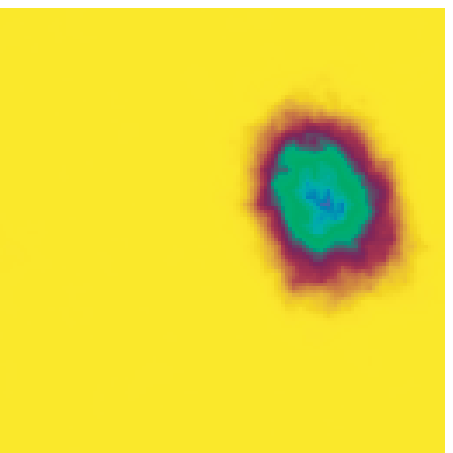} }
\caption{ \label{figyukawa} 
(Color online) Numerical simulation of eqs (\ref{nls}) and (\ref{nls.imp}) for $N=6$ impurity fields in a $64 \, \times 64 \, {\rm units^2}$ periodic plane. Here the initial state is given by  uniform (plus a small fluctuation) wave functions for the condensate and the impurity fields such that  $\int |\psi|^2 d^2{\bm x} = 4096$ and  $\int |\chi_k|^2 d^2{\bm x} = 40.96 $, moreover $m=1$, $\lambda=0.2$, $\gamma=1$ and $\gamma_0=0$. Figure {\it a)} plots the condensate density $|\psi|^2$, and {\it b)} plots $\sum_{k=1}^N |\chi_k|^2 $.  Figures {\it i)} through {\it vi)} each plot a different impurity field $|\chi_k|^2$.  See supplementary material for a movie of crystallization \cite{epaps}.}
\end{center}
\end{figure}


\begin{figure}[htc]
\begin{center}
\centerline{{\it i)} \includegraphics[width=3.5cm]{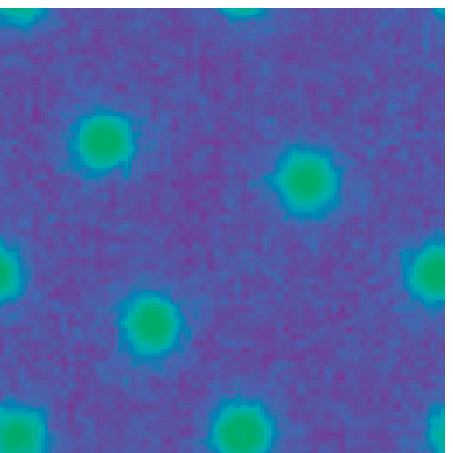} \quad {\it iii)}  \includegraphics[width=3.5cm]{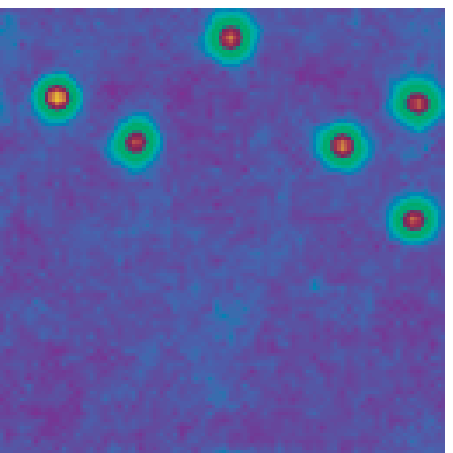}  \quad {\it v)}  \includegraphics[width=3.5cm]{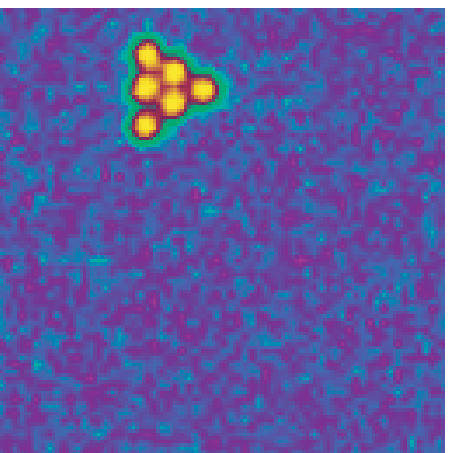}  \quad {\it vii)}  \includegraphics[width=3.5cm]{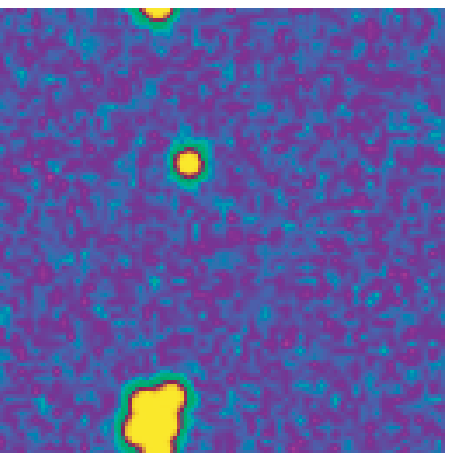} }
\centerline{{\it ii)} \includegraphics[width=3.5cm]{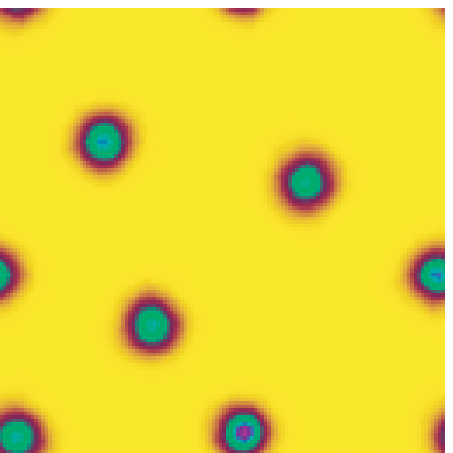} \quad {\it iv)}  \includegraphics[width=3.5cm]{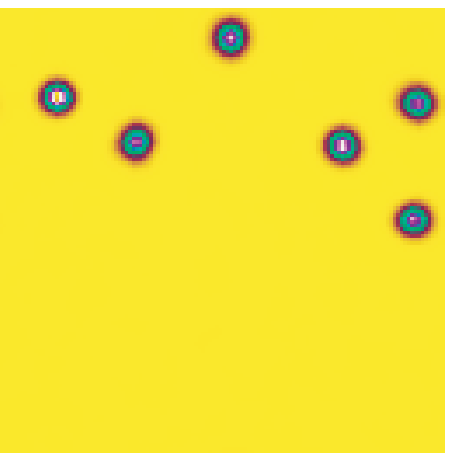}  \quad {\it vi)}  \includegraphics[width=3.5cm]{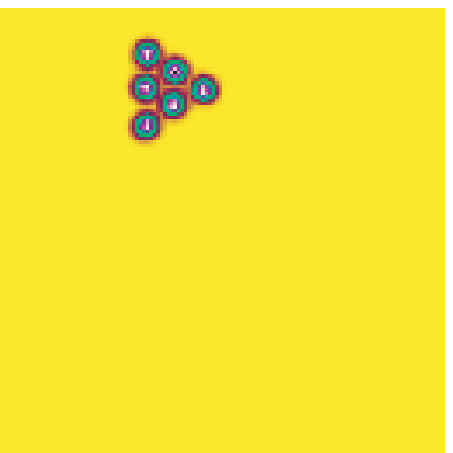}  \quad {\it viii)}  \includegraphics[width=3.5cm]{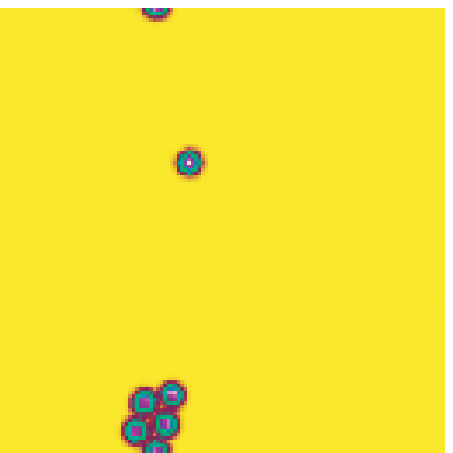} }
\caption{ \label{YukawaPhases} (Color online) Numerical simulation of eqs (\ref{nls}) and (\ref{nls.imp}) for $N=6$ impurity fields in a $64 \, \times 64 \, {\rm units^2}$ periodic plane. Here,  as in previous figure, the initial state is given by  uniform (plus a small fluctuation) wave functions for the condensate and the impurity fields such that  $\int |\psi|^2 d^2{\bm x} = 4096$ and  $\int |\chi_k|^2 d^2{\bm x} = 40.96$, and  $m=1$,  $\gamma=1$ and $\gamma_0=0$. $\lambda$ increases from left to right. The upper row plots the condensate density $|\psi|^2$ while the lower row plots $\sum_{k=1}^N |\chi_k|^2 $.  The corresponding values of $\lambda$ are as follows: $\lambda =0.4$ in {\it i)} and {\it ii)} ; $\lambda =0.5$ in {\it iii)} and {\it iv)}; $\lambda =0.7$ in {\it v)} and {\it vi)}; and $\lambda =1$ in {\it vii)} and {\it viii)}.  The size of the impurity fields decreases as $\lambda$ increases due to the self-interaction term seen in eq. (\ref{selfenergy}). (In the latter case, notice that only five of the impurity fields interact strongly and ultimately form a bound state; the remainder are far enough so that the exponentially weak interaction is negligible.) The case $\lambda =0.2$ is plotted in {\it a)} and {\it b)} of Fig. \ref{figyukawa}. See supplementary material for a movie of crystallization \cite{epaps}.
}
\end{center}
\end{figure}

\subsection{Phase {\it IV}}
In this phase,  $\gamma>\gamma_0$ and $\lambda > \frac{{\sqrt{(N-1)\,\gamma  +{\gamma_0}}}}{{\sqrt{N}}}$. The condensate phase separates from the impurity fields, and the impurity fields also phase separate from each other (unlike in Phase II).  This results in the impurity fields gathering into a ``bubble'' inside of which the condensate density is essentially zero. Within this bubble, the absence of mediating condensate means that the impurity-impurity interaction is no longer Yukawa-type (i.e. $\psi_0\approx 0$ in eq. (\ref{yukawa})), a special case which will be discussed in the next section.  The condensate basically acts as a container by causing a stronger attraction between any impurity field that strays from the bubble and the remaining ensemble of impurity fields.  This inward ``pressure'' from the condensate and the close-range hard-core repulsion bring about crystallization (see Fig. \ref{fig.bubble}).

\begin{figure}[htc]
\begin{center}
\centerline{{\it a)} \includegraphics[width=3cm]{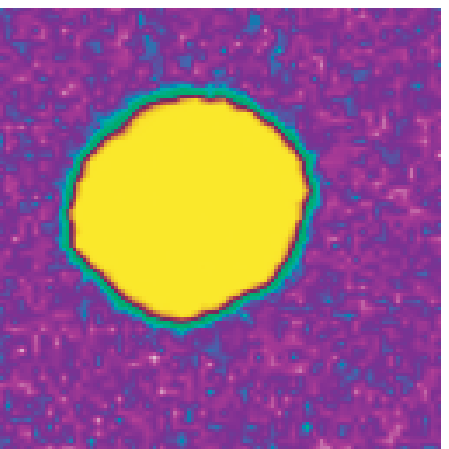} \quad {\it b)}  \includegraphics[width=3cm]{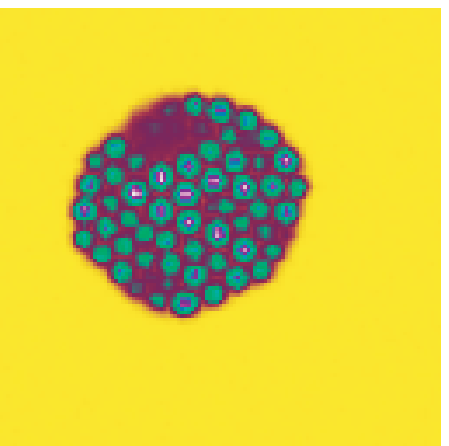}}
\caption{ \label{fig.bubble} 
(Color online) Numerical simulation of eqs (\ref{nls}) and (\ref{nls.imp}) for $N=36$ impurity fields in a $64 \, \times 64 \, {\rm units^2}$ periodic plane. The initial state is given by uniform (plus a small fluctuation) wave functions for the condensate and the impurity fields such that $\int |\psi|^2 d^2{\bm x} = 4096$ and  $\int |\chi_k|^2 d^2{\bm x} = 40.96 $, $\lambda=2$, $\gamma=1.5$ and $\gamma_0=0$. Plot {\it a)} depicts the condensate density $|\psi|^2$ and plot {\it b)} depicts $\sum_{k=1}^N |\chi_k|^2 $.  }
\end{center}
\end{figure}

\section{Analytical treatment of Phase {\it IV}: The case $\lambda=0$}\label{LambaEq0}

In phase {\it IV} described above, the system is at a special limit when the condensate is not coupled to the impurity fields. In such a case, the induced attractive interaction described earlier is no longer valid.  Therefore, the case $\lambda =0$ deserves special attention. The condensate can be described by the pure nonlinear Schr\"odinger equation, which is quite well understood. The impurity fields, however, evolve and we shall now consider this evolution.  Impurity fields that overlap store potential energy and will tend to repel each other; more so as $\gamma$ increases.  If the impurity fields $|\chi_k|^2$ are localized about ${\bm R}_k$ with a size $\delta$ then, in the limit $\gamma\gg \gamma_0$, the potential energy is dominated by $$\frac{\gamma}{2} \sum_{i\neq k}  \int |\chi_i({\bm x}) |^2 |\chi_k({\bm x})|^2  d^D{\bm x}.$$  Minimizing the overlap will minimize the potential energy.   In some sense, the energy may be approximated by the superposition of two-body interactions ${\mathcal E}= \frac{\gamma}{2} \sum_{i\neq k}  U(|{\bm R}_i - {\bm R}_k |  )$, with $U(d)$ repulsive and Gaussian (see below). However, in a finite box there is a limit to how far away from each other the impurities can move.  A kind of close-packing argument may be invoked to find the crystal structure that minimizes the energy of the impurity ensemble, sustained only by the external pressure from the condensate at the boundaries. 

\begin{figure}[htc]
\begin{center}
\centerline{\includegraphics[width=4cm]{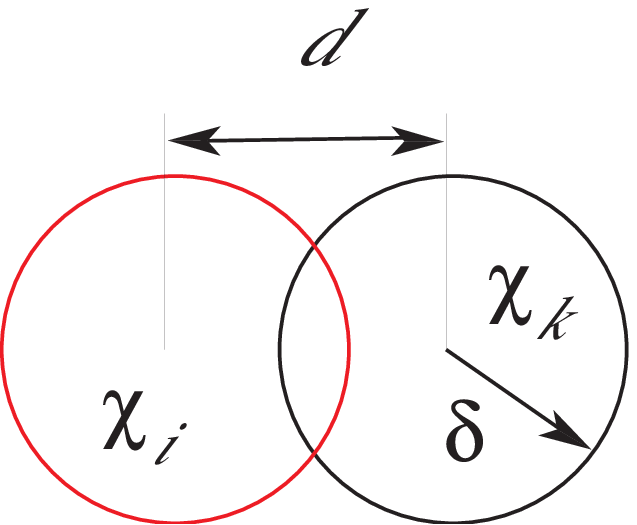}  }
\caption{ \label{Imps} 
(Color online) Sketch of the minimum energy (overlap) configuration for two impurity fields.  The size of the localized fields is $\delta$ and they are separated by $d$. }
\end{center}
\end{figure}

The periodic case, i.e. when all $N$ impurity fields have the same number of particles $n_{*}$, provides an interesting, solvable example. One can use the minimization approach to determine the size $\delta$ as a function of the large parameter $\gamma$. Only the nearest neighbor(s) affect the interaction energy. To begin, let us assume that the impurity field is a compact function (i.e., it vanishes exactly outside the ball of radius $\delta$) and that there is no overlap.  In this case, one can calculate the minimum of the total energy.  We shall assume $\gamma_0=0$ in the following and note that there is no overlap, which means no interaction energy  $\frac{\gamma}{2} \sum_{i\neq k}  \int |\chi_i({\bm x}) |^2 |\chi_k({\bm x})|^2  d^D{\bm x}$.  The total energy can be written as

\begin{eqnarray}
H = \sum_{k=1}^{N}  \int  \frac{1}{2m} | \nabla \chi_k |^2    d {\bm x}.\label{hamilton}
\end{eqnarray}
 The Euler-Lagrange conditions for an extreme of this energy leads to the Hemholtz equation
\begin{eqnarray}-\frac{1}{2m} \Delta \chi_k = \varepsilon \chi_k
\label{hemholtz}
\end{eqnarray}
inside a ball $V_\delta({\bm R}_k)$ and with a Dirichlet boundary condition on  
$\partial V_\delta({\bm R}_k) $ \cite{amandine}.

From here, one can proceed using perturbation to determine the effect of a slight overlap.  The variational parameter $\delta$ will determine the optimal configuration  that balances the kinetic energy of a pulse and the interaction energy due to a small overlap.

As an example, let us consider the one-dimensional case in a periodic domain of length $L$. Let $\chi_s (x)=  \sqrt{n_{*}} f( x- s d) $  with $d = L/N $, the ground state.  The energy is dependent on the number of impurity fields (which is the same as the number of sites) such that

\begin{eqnarray}E[f]& =&  \frac{N \, n_*}{2}  \int \left[ \frac{1}{m} |f'(x)|^2  + \gamma \, n_*  f(x)^2  \sum_{n\neq 0} f(x- n d)^2  \right]dx 
\end{eqnarray}


The energy may be estimated explicitly by using 
\begin{eqnarray}
f_\delta(z) =\frac{1}{\sqrt{\delta}}  \cos({\pi z}/{(2\delta)} ),
\label{Ansatzfdelta}
\end{eqnarray}
the solution of the 1D Hemholtz equation (\ref{hemholtz}).
This yields

\begin{eqnarray}E& =& N n_*   \frac{1}{2}   \int_{-\delta}^{\delta} [  \frac{1}{m}|f_\delta'(x)|^2  + \gamma  n_*  f_\delta(x)^2 (f_\delta(x- d)^2  + f_\delta(x- d)^2 ) ] dx \nonumber\\
&=&  N  n_* \left[ \frac{{\pi }^2}{ 8 \, m\,{\delta }^2} + 
      \frac{n_*\,\gamma}{ 8\,\pi \,{\delta }^2} \,\left(  
         3\,\delta \,\sin (\pi d/\delta  )  -\pi \,
            \left( d - 2\,\delta  \right) \,
            \left( 2 + \cos(\pi d/\delta  )
              \right)  
         \right)  \right]  .\nonumber
\end{eqnarray}
The dimensionless parameter $\zeta = \frac{d\,\pi }{\delta }$ that minimizes this energy satisfies the implicit equation 
$${ m \, n_*\,\gamma \, d} =\frac{2\,{\pi }^2\,\zeta }
 {\left( -4\,\pi  + 4\,\zeta  - 
  \left( 2\,\pi  + \zeta  \right) \,\cos (\zeta ) + 
  \left( -3 + 2\,\pi \,\zeta  - {\zeta }^2 \right) \,
  \sin (\zeta) \right) },$$ 
which in the large $\gamma$ limit ($\zeta \rightarrow 2 \pi$) gives $$\delta \approx  \frac{d}{2}\left( 1 + \frac{1}{\sqrt{\pi }}\left( \frac{3}{2} \right) ^ {\frac{1}{4}} (m n_* \gamma d )^{-\frac{1}{4}} + \dots \right).$$     
 Thus, as expected, $\delta \rightarrow d/2 $ as $\gamma \rightarrow\infty$.
 In a similar approximation, the energy becomes
 \begin{equation}
 E =   N n_*   \frac{\pi^2}{2md^2}   \left[ 1 -   \frac{2}{\sqrt{\pi}} \left(\frac{3}{2 d m n_* \gamma}\right)^{1/4}  \right].
 \label{energy2}
 \end{equation}
 
The compact solution $f_\delta(z)$ is only an approximation since the real ground state does not vanish entirely at $x = \pm \delta$ even though the impurity field is exponentially small.  The nonlinear structure of the boundary layer may be approximated by the ordinary differential equation that satisfies the function $f$ :
$$- \frac{1}{2m} f'' + \gamma n_k  f(x) ( f(x-d)^2 + f(x+d)^2 )  = \varepsilon f(x).$$
We are interested in the region where $x\approx\delta$, so one may neglect the $ f(x+d)^2 $, which is an exponentially small term.  We approximate the second term by  $ f_\delta(x-d)^2 $  from eq. (\ref{Ansatzfdelta}), which is a given and known function.


The remaining linear ordinary differential equation may be solved in the WKB (large $\gamma$) limit, $f(x) \sim e^{-\sqrt{\gamma} S(x)}$, so that 
$\frac{1}{2m} S'(x)^2 \approx  n_k  f_\delta(x-d)^2$. Therefore,
$$S(x) \approx \frac{2 }{\pi} \sqrt{2\, m\, n_k \, \delta} \sin(\frac{\pi (x-d) }{2\delta}).$$

Finally, near $x\approx \delta\approx d/2$ (as shown previously $\delta \approx d/2$), we find

$$ f(x) \sim e^{ -\pi \sqrt{ \gamma m\, n_k} (x-d/2)^2}.$$ 

The domination of the interaction $U(d)$ by this Gaussian tail shows that impurity fields can be free of each other's influence with sufficient separation.  Also, when the separation is sufficiently small, the hard-core repulsion term comes into play, making the impurities impenetrable  (see Fig. \ref{CaseLambdaEqZero}).

\begin{figure}[htc]
\begin{center}
\centerline{{\it a)} \includegraphics[width=5cm]{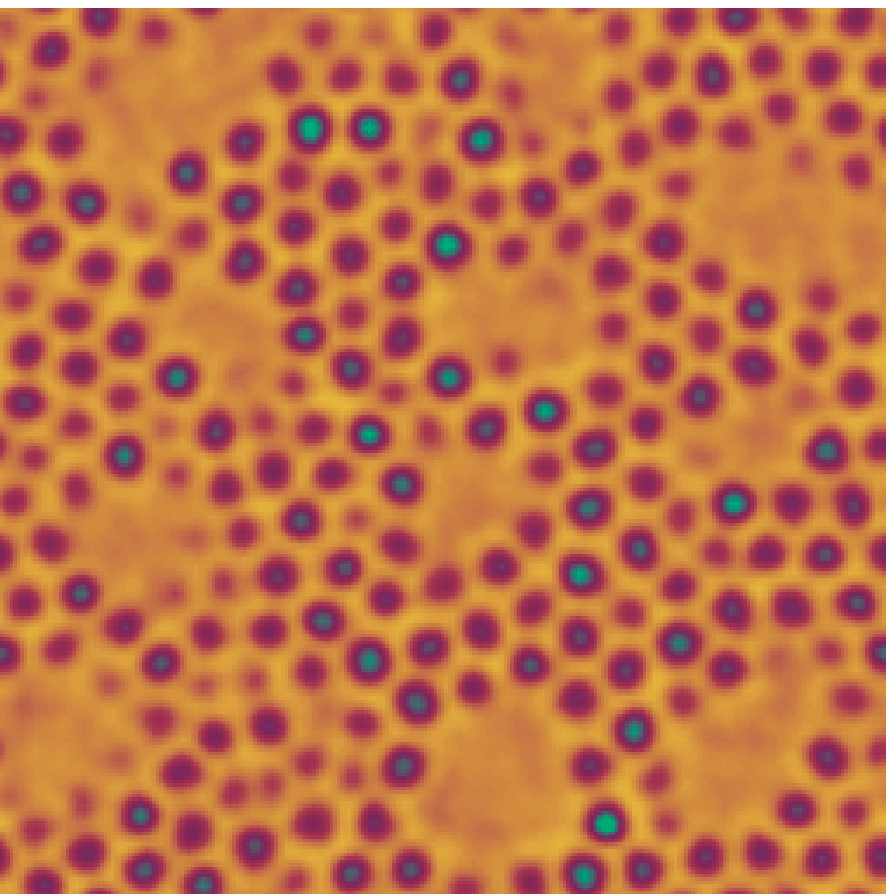} \quad {\it i)}  \includegraphics[width=5cm]{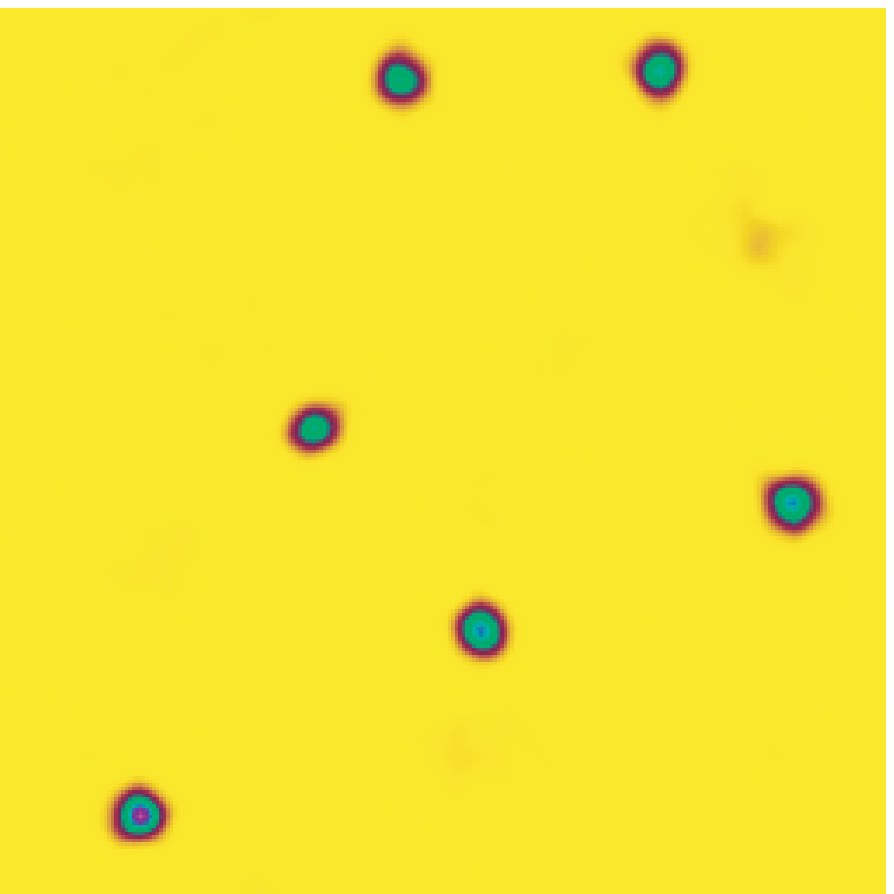} \quad {\it ii)}  \includegraphics[width=5cm]{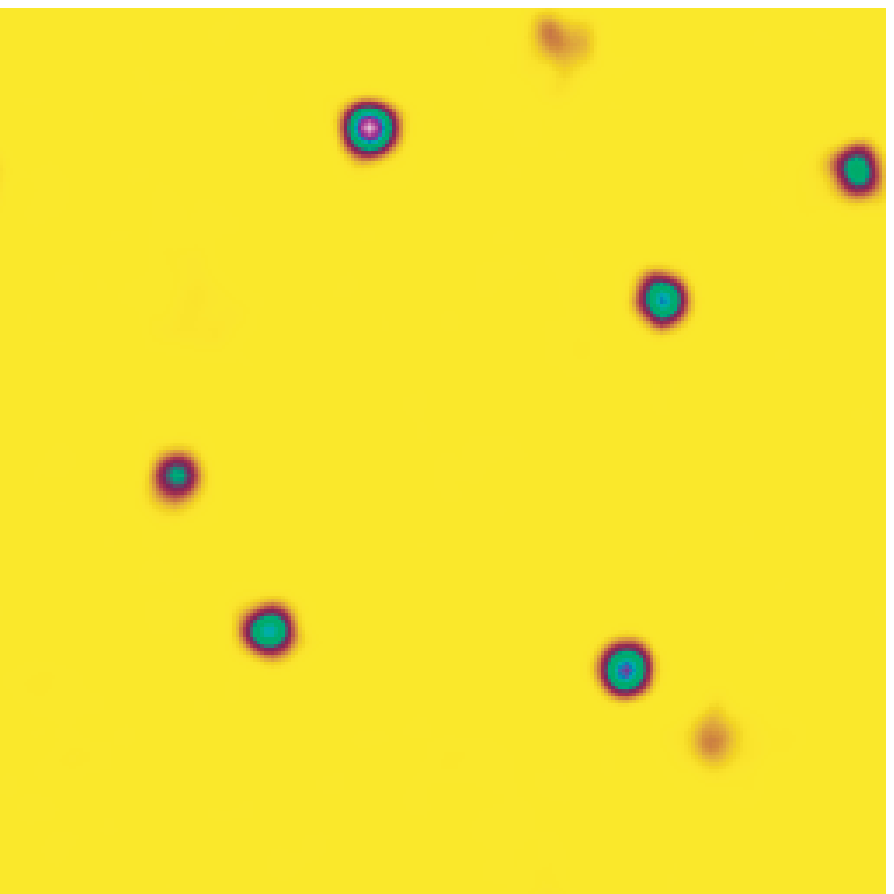}}
\caption{ \label{CaseLambdaEqZero} 
 (Color online) Numerical simulation of eq. (\ref{nls.imp}) for $N=36$ impurity fields with $\lambda=0$ in a $128 \, \times 128 \, {\rm units^2}$ periodic plane. Here the initial condition is an uniform miscible state plus small fluctuations, the  number of particles are given by  $\int |\chi_k|^2 d^2{\bm x} = 200 $,  $\gamma=1$, and $\gamma_0=0$. The plots represent the instant $t=1700 \, units$ : {\it a)} shows $\sum_{k=1}^N |\chi_k|^2 $; and {\it i)} and {\it ii)}  plot two of the 36 distinct impurity fields.  Note that the system evolves through  a Hamiltonian (conservative) dynamics though the relaxation to an ``equilibrium" takes a long time. Indeed, at the moment depicted in the plots above, the system has yet to reach such an equilibrium, so each impurity field contains many distinct localized structures.  We remind the reader that while the analysis done above is in $1D$, the numerical simulations are in $2D$.}
\end{center}
\end{figure}

 In conclusion, a crystal phase is possible whenever the condensate presence is negligible and the impurity fields are immiscible. The crystal is composed of distinct localized impurity fields which are repulsive and  impenetrable. Therefore, the crystal lattice is a result of the effect of  the boundaries: if $V$ is the volume then the mean separation distance would be  $\sim (V/N)^{1/D}$ which determines the localized structure size. This crystal will not exist without the presence of boundaries or a containing pressure. The condensate acts naturally as this pressure as seen in Fig. \ref{fig.bubble}. By construction, it seems plausible that these crystallites (like those of Fig.\ref{SimpleDiagram}-{\it IV}, Fig. \ref{YukawaPhases}-{\it viii)}, or Fig. \ref{fig.bubble}) possess a surface energy. In general, the energy of the system in phase {\it IV} is made up of a bulk contribution from the condensate component ($\psi\approx \psi_0$ and  $\chi_k\approx 0$) set to zero in the energy given in eq. (\ref{free})), the energy of the crystal phase ($\psi\approx 0$ and $\chi_k$ a crystal structure built in the way explained in this section), and an interphase energy, which we have not computed. However, by general arguments one can expect that the ratio of this interphase energy to the bulk energy  from the crystal gives a critical radius of crystallite nucleation. It also says that the transition between phases {\it III} and {\it IV} is of first order. However, we were unable to observe a  hysteresis effect (in Fig. \ref{Hysteresis}) when the crystallite is formed and one decreases $\lambda$ through the boundary.  This implies the interface energy is weak.

 \begin{figure}[htc]
\begin{center}
\centerline{{\it a)} \includegraphics[width=3cm]{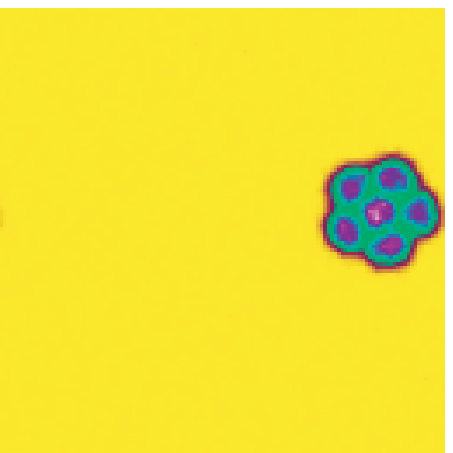} \quad {\it b)} \includegraphics[width=3cm]{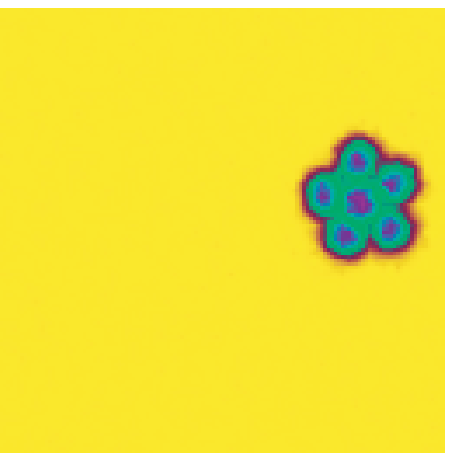} }
\centerline{{\it c)} \includegraphics[width=3cm]{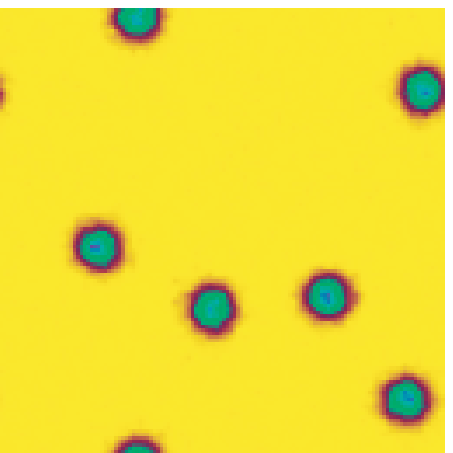} \quad {\it d)} \includegraphics[width=3cm]{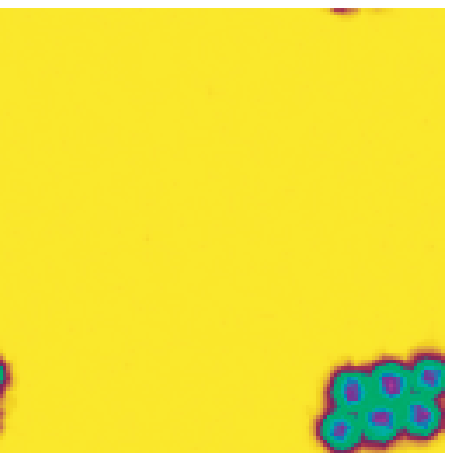} }
\caption{ \label{Hysteresis} (Color online) Numerical simulation of eqns. (\ref{nls}) and (\ref{nls.imp}) for $(N=6)$ impurity fields with $\gamma_0=1$ in a $64 \, \times 64 \, {\rm units^2}$ periodic plane where $\int |\psi|^2 d^2{\bm x} = 4096$ and  $\int |\chi_k|^2 d^2{\bm x} = 40.96 $. We begin with a crystallite initial state that comes from the same simulation as in Fig. \ref{SimpleDiagram}-phase {\it IV}, and subsequently we slowly decrease $\lambda $ from $\lambda =2 $ to below  the critical value $\frac{\sqrt{(N-1)\,\gamma  +\gamma_0} }{\sqrt{N}}= \sqrt {11/6} \approx 1.354$. The plots represent $\sum_k |\chi_k|^2 $ and the colormap is the same for all four images. The parameters are $\lambda=1.5$ at the stage shown in  {\it a)}, $\lambda=1.3$ in {\it b)}, $\lambda=1.25$ in  {\it c)}, and then we increase back to $\lambda=1.3$ in {\it d)}.  We do not observe a hysteresis effect.}
\end{center}
\end{figure}


\section{Nonclassical rotational inertia}\label{SecNCRI}
In this section, we will investigate the crystaline phases in phases {\it III} and {\it IV} retain their superfluid behavior, thus exhibiting properties reminiscent of a supersolid \cite{ss1,ss2,ss3}.  Inspired by Leggett's seminal work \cite{leggett}, consider a condensate-impurity system confined within an annulus of total length $L$ and total volume  $V= S L$. Consider this system to be rotating uniformly about the annulus's primary axis of rotation, $\hat e_1$ (the direction associated with the coordinate $x_1$ which spans [0,$L$]), a rotation induced by the boundary condition $\psi(x_1=L) = \psi(x_1=0) e^{i \alpha_{0}}$ where $\alpha_{0} = \frac{m_{0} \omega L^2}{\hbar} $ is dimensionless, and $\chi_k(x_1=L) = \chi_k(x_1=0) e^{i \alpha_{imp}}$ where $\alpha_{imp} = \frac{m \omega L^2}{\hbar} $. Then the system will possess an energy (for $\alpha_0 \ll 1$ and  $\alpha_{imp} \ll 1$)
$${\mathcal E} = {\mathcal E} _0 +  \Delta {\mathcal E} $$
where ${\mathcal E} _0 $ is the ground state energy and $  \Delta {\mathcal E} =\frac{1}{2} {I}^{eff}\omega^2$, $ {I}^{eff} $ being the effective and observable (measurable) moment of inertia tensor around the $\hat {\bm e}_1$ axis. 

The relative deviations of this tensor with respect to the rigid body rotation, $I^{RB}=\rho L^2 V$ where $\rho$ is the total mass density, is called the nonclassical rotational inertia (NCRI) fraction of the annulus.  NCRI is seen as a signature of a superfluid response of a system. 

Let $\psi^0$ and $\chi_k^0$ be the ground state real wave functions of the Bose field and the impurity  fields respectively. Under rotation, the phases of these fields are no longer uniform,  so the rotating wave functions are of the form $\psi^0({\bm x}) e^{i \phi} $ and $ \chi_k^0({\bm x})  e^{i\phi_k} $ (the only change comes from the phase to the lowest order)   and the increase of energy is 

$$\Delta {\mathcal E}  =   \frac{1}{2}  \int \left(  \rho_0({\bm x} )\, | \nabla \phi |^2  +  \sum_{k=1}^{N}  \frac{1}{m}  \rho_k({\bm x} ) | \nabla \phi_k |^2   \right)  d {\bm x}$$
where  $ \rho_0({\bm x} )  = |  \psi^0({\bm x}) |^2 $ and $ \rho_k({\bm x} )  = |  \chi_k^0({\bm x}) |^2  $ are the non-uniform density of the respective ground states, and the phases $\phi$ and $\phi_k$ satisfy the boundary conditions explained below. That $\Delta {\mathcal E}$ is a quadratic form in $\omega$, i.e. in $\alpha$, is quite evident since $\phi $ and $\phi_k$ are proportional to $\alpha_{0}$ and $\alpha_{imp}$ because of the boundary conditions (see below). The explicit form of ${I}^{eff}$ and the prefactor, on the other hand, need longer consideration.

The minimization of $\Delta {\mathcal E}$ leads to the equations and boundary conditions (recalling the coordinates other than $x_1$ are periodic)
\begin{eqnarray}
{\bm \nabla} \cdot (\rho_0({\bm x} )\,  \nabla \phi ({\bm x} ) )  =0 \quad \in L^D& &  \& \quad  \phi(x_1=L) = \phi(x_1=0) +  \alpha_{0}   \quad\label{cont1}\\
{\bm \nabla} \cdot (\rho_k({\bm x} )\,  \nabla \phi_k({\bm x} ) )  =0  \quad \in L^D & & \& \quad  \phi_k(x_1=L) = \phi_k(x_1=0) +  \alpha_{imp} .
\label{cont2}
\end{eqnarray}

Eqs (\ref{cont1}) and (\ref{cont2}) may be solved using the method called homogenization \cite{homo}.  This method splits cleanly the large (system size $L$) and small (impurity size) scales and provides effective average quantities. Consider for example eq. (\ref{cont1}), taking $ \phi ({\bm x} )= \frac{\alpha_{0}}{L} x_1 +  \tilde \phi ({\bm x} )$  where $ \tilde\phi ({\bm x} )$ is periodic in $L^D$. The periodic function $\tilde\phi ({\bm x} ) \equiv  \alpha_{0} K({\bm x} ) $ satisfies 
$$ {\bm \nabla} \cdot (\rho_0({\bm x} )\,  \nabla K({\bm x} ) )  +  \hat {\bm e}_1 \cdot  {\bm \nabla}  \rho_0({\bm x} )=0$$
and, after substituting this back into $\Delta {\mathcal E}$, one finally obtains $$\Delta {\mathcal E}  =   \frac{\hbar^2}{2m_{0}^2}  \varrho^{ss}_{0} \frac{\alpha_{0}^2  }{L^2} V $$
where 
\begin{equation} \varrho^{ss}_{0}=  \frac{m_{0} }{V} \int_{V}  \rho_0({\bm x} )( 1 -  ({\bm  \nabla }K)^2 ) d^D {\bm x} \label{rhoss}
\end{equation}
and has units of a mass density. In fact, it is the superfluid density. Notice that by replacing $\alpha_{0}$ one can identify the effective moment of inertia $I^{eff}  =   \varrho^{ss} {L^2}  V $, which is the moment of inertia of an annulus with effective density mass $ \varrho^{ss}$.
It is not possible to obtain a closed expression for $\varrho^{ss} $ in terms of the local density $ \rho_0({\bm x} )$  ($K$ implicity depends on $\rho_0$). However, it is possible to show that $I^{eff} \leq I^{RS}$, and hence the system displays a NCRI. In one-dimensional space, eq. (\ref{cont1}) may be solved exactly, with a direct calculation leading to the  formula
 \begin{equation} \varrho^{ss}_{0}= m_{0}  \left(\frac{1}{V} \int_{V}  \frac{1}{\rho_0({x} ) } d {x} \right)^{-1}, \label{rhossleggett}
\end{equation}
due to Leggett \cite{leggett}.

Finally, to take into account the contributions from the condensate and impurity field, one need simply add the corresponding superfluid densities; and because the impurity fields are localized, their wave functions decay fast in space so the contribution to the superfluid density from the impurity fields becomes negligible.  Therefore, perhaps unsurprisingly, the superfluid density is dominated by the condensate density.

\section{Discussion}

In this paper, we analyzed how the condensate-mediated interactions
between self localized impurity fields can lead to pattern formation.
We restricted ourselves to the case where all impurity-impurity
interactions are governed by the same coupling constant.  In ultracold
trapped atomic gas experiments, impurity fields may be generated by
populating different atomic levels and/or by making use of different
atomic species/isotopes (for review of current experimental progress  on multicomponent condensates see section IX of \cite{nonlinearbook} ).  In these cases, unlike pairs of impurity
field would have different coupling constants.  In this situation, we
would expect there to be phases similar to those described in
section~\ref{Crystal}, but that the broken translational symmetry of
the Hamiltonian will lead to nonperiodic structures, rather than
periodic order.

\appendix

\section{Dynamic Instabilities of the Miscible State:  The Bogoliubov Spectra and Finite Size Effects}\label{AppA}
In this Appendix we work out the dynamic instability conditions of the uniform miscible state, and show that these conditions are important when we take into account the fact that the system is of finite size.  Consider the case of uniform ground states, $\psi = \psi_0 e^{- i \Omega_0 t}$ where $\Omega_0 = |\psi_0|^2 + \lambda \sum_{j=1}^N |\chi^0_j |^2$, and $\chi_j  =  \chi^0_j  e^{- i \Omega_j t}$ where $\Omega_j = \lambda |\psi_0|^2 + \gamma_0 |\chi^0_j |^2 + \gamma  \sum_{k\neq  j}^N |\chi^0_k |^2$.  The $N+1$ modes with wavenumber $k$ around this ground state would obey the $N+1$ Bogoliubov dispersion relations   \cite{robertsueda} , $$ \omega^{(s)}_k = \sqrt{ e^{(s)}_k }  $$
where $s$ runs from 1 to $N+1$ and $e^{(s)}$ is the $s^{th}$ eigenvalue of the matrix 

\begin{equation}
{\mathcal M}_B =  \frac{k^4}{4} \left( \begin{array}{cccccc}
1 &0 &0 & \cdots & 0 &0 \\
0 & 1/m  &0 & \cdots &0  &0  \\
0&0  &1/m   & \ddots &  & 0 \\
\vdots & \vdots & \ddots & \ddots& \ddots& \vdots \\
0  & \vdots &   &  \ddots &1/m & 0 \\
0 & 0& \cdots  & \cdots &0 & 1/m  \\
			\end{array}\right)
 +k^2  \left( \begin{array}{cccccc}
 |\psi_0|^2 &\lambda  |\psi_0|^2 & \lambda   |\psi_0|^2 & \cdots &  \lambda  |\psi_0|^2 & \lambda  |\psi_0|^2 \\
\lambda  |\chi^0_1|^2/m & \gamma_0  |\chi^0_1|^2/m  &\gamma  |\chi^0_1|^2/m  & \cdots &  \gamma  |\chi^0_1|^2 /m & \gamma  |\chi^0_1|^2/m  \\
\lambda  |\chi^0_2|^2/m & \gamma  |\chi^0_2|^2/m  &\gamma_0  |\chi^0_2|^2 /m  & \ddots &  & \gamma  |\chi^0_2|^2/m \\
\vdots & \vdots & \ddots & \ddots& \ddots& \vdots \\
\lambda |\chi^0_{N-1}|^2 /m & \vdots &   &  \ddots & \gamma_0  |\chi^0_{N-1}|^2/m& \gamma  |\chi^0_{N-1}|^2/m \\
\lambda |\chi^0_N|^2/m & \gamma|\chi^0_N|^2 /m & \cdots  & \cdots & \gamma |\chi^0_N|^2 /m&  \gamma_0 |\chi^0_N|^2 /m \\
			\end{array}\right).\label{MatrixMB}
\end{equation}

From the construction of the matrix ${\mathcal M}_B$ it is easy to show that if ${\mathcal M}$, eq. (\ref{MatrixM}), is positive semidefinite then ${\mathcal M}_B$ is also positive semidefinite. To do so, we note first that the sum of two positive matrices is a positive matrix and that the first matrix on the RHS of eq. (\ref{MatrixMB}) is naturally positive; then, for the second matrix on the RHS of eq. (\ref{MatrixMB}), we recall that a matrix is positive semidefinite if the determinants of all upper left submatrices are non-negative and note that the determinants of all the upper left submatrices  of  ${\mathcal M}_B$ differ from the corresponding submatrices of ${\mathcal M}$ only by a factor of $ |\psi_0|^2 $ or $|\chi^0_j|^2 $, both of which are positive.   
 
The eigenvalues of this second matrix play an important role in the dynamics because they are they charaterize the instabilities at  long wavelengths. In the convenient case where all $|\chi^0_j |^2 = |\chi^0|^2 $, the eigenvalues are $(\gamma_0-\gamma) |\chi^0|^2 /m $, which is $N-1$ degenerate, and 
\begin{equation}
\frac{1}{2} \left(    |\psi_0|^2 + ((N-1) \gamma +\gamma_0)  |\chi^0|^2/m   \pm \sqrt{  (-|\psi_0|^2 + (   (N-1) \gamma+\gamma_0  )  |\chi^0|^2/m ) ^2 + 4 N  \lambda^2 |\psi_0|^2  |\chi^0|^2/m } \right).
\label{BogSpectra}
\end{equation}
We can see that a modulational instability occurs if either $\gamma>\gamma_0$ or $\lambda > \frac{{\sqrt{(N-1)\,\gamma  +{\gamma_0}}}}{{\sqrt{N}}}$, which are the same as the energetic instability conditions described in section \ref{InstabilityCriteriaMatrixM}.  
      
Therefore, the dynamic instability conditions (from the Bogoliubov analysis) and the energetic instability conditions are identical in the infinite system. Note that the values of the densities of the condensate  $|\psi_0|^2 $ and impurity fields $ |\chi^0_k|^2$ are absent in these conditions since they only play a role in the time and length scales involved.  

However, in a finite system | take for instance a periodic box of size $L$ these transition lines are shifted because the lowest mode (where the instability typically arises) is nonzero.  Indeed, the growth instability rate becomes positive if one of the eigen-frequencies $\omega^2_k  < 0 $.  That is, the critical line is shifted by $ \frac{\pi^2}{L^2} $. So, assuming $|\chi^0_j|^2=|\chi^0|^2$ as before, the degenerate eigenvalue $(\gamma_0-\gamma) |\chi^0|^2 $ (responsible for the impurity immiscibility)  becomes negative when
$\gamma_0 + \frac{\pi^2}{L^2 |\chi^0|^2 } < \gamma $ (instead of $\gamma_0 < \gamma $). 


 One can also show the instability condition responsible for the condensate-impurity immiscibility is shifted as can been seen in the large $L$ limit:

\begin{equation}
\lambda  > \frac{{\sqrt{(N-1)\,\gamma  +{\gamma_0}}}}{{\sqrt{N}}}  + \frac{|\psi_0|^2 + ((N-1)\,\gamma  +{\gamma_0}) |\chi^0|^2 }{2|\chi^0|^2 |\psi_0|^2 \sqrt{N((N-1)\,\gamma  +{\gamma_0})}}  \frac {\pi^2}{L^2} +{ \mathcal O}(1/L^4)
\label{UedaCondition}
\end{equation}

where the latter approximation is valid in the limit $  |\chi^0|^2 \ll |\psi_0|^2$. One notices that this second condition is shifted by a smaller amount as the number of impurities $N$ becomes larger because of the $\sqrt{N(N-1)}$ in the denominator.

\section{Numerical tools}
\label{AppB}

 We used a Gauss-Seidel Crank-Nicholson finite difference method to integrate the Hamiltonian coupled equations (\ref{nls},\ref{nls.imp}). The scheme is as follows: let $U(t)= \{ \psi(t), \chi_1(t) \dots \chi_N(t)  \} $ be the fields at time $t$, and let us write eqs (\ref{nls},\ref{nls.imp}) as
$  \frac{\partial U}{\partial t}  = F[U(t)] $ where $F[U]$ is a nonlinear Hermitian operator whose definition is clear by comparison with eqs (\ref{nls},\ref{nls.imp}). Thus, one time step  from $t$ to $t+dt$ is
$$U(t+dt) - \frac{dt}{2} F[U(t+dt)] = U(t) + \frac{dt}{2}  F[U(t)].$$
For small $dt$, this (classical) numerical schema yields $\psi(t+dt)$, exact to at least the second order in $dt$ and has the advantage that, formally, the norm and energy are exactly conserved. However, the price is that $U(t+dt)$ involves calculation from a complex nonlinear equation. Next, to obtain $U(t+dt)$, one iterates the mapping $ u_{n+1}= (U(t) + \frac{dt}{2}  F[U(t)] ) +  \frac{dt}{2} F[u_n] $ where $  (U(t) + \frac{dt}{2}  F[U(t)] )$ is constant, $u_{n=0}=U(t)$, and $u_{n\rightarrow\infty} =  U(t+dt)$. Convergence is expected after a moderate number of iterations, 6 in practice, using a time step $dt = 0.01$ units. Under this condition, the norm and the energy of the solution are conserved in time, deviating by less than one part per $10^8$ per time step and less than one part per $10^7$ per unit time.

For most of the figures, the numerical simulation of eq. (\ref{nls.imp}) was performed in a $64 \, \times 64 \, {\rm units^2}$ periodic plane for $N=6$ impurity fields.  Unless otherwise stated, the  simulations typically use $\int |\psi|^2 d^2{\bm x} = 4096 $ and $\int |\chi_k|^2 d^2{\bm x} = 40.96 $.

\begin{acknowledgements}

SR would like to thank the Center of Nonlinear Studies at Los Alamos
National Laboratory for their hospitality where part of this work was
done, and the Agence Nationale de la Recherche  ANR-08-SYSC-004
(France).  Similarly, DCR is grateful to the Universidad de Chile and Ecole
Normale Superieure for facilitating collaboration on this work through
their hospitality.

\end{acknowledgements}

\end{document}